\documentclass[manuscript]{aastex}

\slugcomment{Not to appear in Nonlearned J., 45.}

\shorttitle{HD189733b}
\shortauthors{Morello et al.}

\begin{document}


\title{A new look at Spitzer primary transit observations of the exoplanet HD189733b}


\author{G. Morello\altaffilmark{1}, I. P. Waldmann, G. Tinetti}
\affil{Department of Physics \& Astronomy, University College London, Gower Street, WC1E6BT, UK}
\email{giuseppe.morello.11@ucl.ac.uk}
\altaffiltext{1}{Dipartimento di Fisica, Universit\`a degli Studi di Palermo, via Archirafi, 90123, Italy}

\author{G. Peres}
\affil{Dipartimento di Fisica e Chimica (previously Dipartimento di Fisica), Specola Universitaria, Universit\`a degli Studi di Palermo, Piazza del Parlamento 1, 90123, Italy}

\author{G. Micela}
\affil{INAF -  Osservatorio Astronomico di Palermo, Piazza del Parlamento 1, 90134, Italy}

\and

\author{I. D. Howarth}
\affil{Department of Physics \& Astronomy, University College London, Gower Street, WC1E6BT, UK}







\begin{abstract}
Blind source separation techniques are used to reanalyse two exoplanetary transit lightcurves of the exoplanet HD189733b recorded with the IR camera IRAC on board the Spitzer Space Telescope at 3.6$\mu$m during the ``cold'' era. These observations, together with observations at other IR wavelengths, are crucial to characterise the atmosphere of the planet HD189733b. Previous analyses of the same datasets reported discrepant results, hence the necessity of the reanalyses.
The method we used here is based on the Independent Component Analysis (ICA) statistical technique, which ensures a high degree of objectivity. The use of ICA to detrend single photometric observations in a self-consistent way is novel in the literature.
The advantage of our reanalyses over previous work is that we do not have to make any assumptions on the structure of the unknown instrumental systematics. Such ``admission of ignorance'' may result in larger error bars than reported in the literature, up to a factor $1.6$. This is a worthwhile trade-off for much higher objectivity, necessary for trustworthy claims.
Our main results are (1) improved and robust values of orbital and stellar parameters, (2) new measurements of the transit depths at 3.6$\mu$m, (3) consistency between the parameters estimated from the two observations, (4) repeatability of the measurement within the photometric level of $\sim 2 \times 10^{-4}$ in the IR, (5) no evidence of stellar variability at the same photometric level within 1 year.
\end{abstract}





\section{Introduction}

Observations of exoplanetary transits are a powerful tool to investigate the nature of planets around other stars. Transits are revealed through periodic drops in the apparent stellar brightness, due to the interposition of a planet between the star and the observer. The shape of an exoplanetary transit lightcurve depends on the geometry of the star-planet-observer system and the spatial distribution of the stellar emission at the wavelength at which observations are taken \citep{ma02}. By solving the inverse problem, it is possible to characterise fully the planet's orbit (Period, $P$; semimajor axis, $a$; inclination, $i$; eccentricity, $e$; and argument of periastron, $\omega$), and to measure its radius, $r_p$ \citep{sea03, kip08, ma02}. Knowledge of the inclination enables determination of the mass of the planet, $m_p$, if $m_p \sin{i}$ is known from radial-velocity measurements.

Multiwavelength transit observations can be used to characterise the atmospheres of exoplanets, through differences in the transit depths, typically at the level of one part in $\sim 10^{4}$ in stellar flux for giant planets \citep{brown01, sea00, tin07b}. For this purpose, the diagnostic parameter is the wavelength-dependent factor $p=r_p/R_s$, i.e. the ratio between the planetary and the stellar radii (or its square, related to the transit depth).

The exoplanet HD189733b is one of the most extensively studied hot Jupiters: the brightness of its star allows spectroscopic characterisation of the planet's atmosphere.

The 3.6$\mu$m transit depth for the exoplanet HD189733b has been debated in the literature. Different analyses of the same dataset, including two simultaneous Spitzer/IRAC observations at  3.6$\mu$m and  5.8$\mu$m, have been used to infer the presence of water vapour in the atmosphere of HD189733b \citep{bea08, tin07}, or to reject this hypothesis \citep{des09}. Another analysis of this dataset is reported by \cite{ehr07}, but we do not comment further their results, as they were not conclusive, because of the very large error bars. \cite{des11} reported the analysis of a second Spitzer/IRAC dataset at  3.6$\mu$m using the same techniques. Their new estimates of the planet's parameters were significantly different from those reported previously by the same authors \citep{des09}; the discrepancies were attributed by the authors to variations in the star.

Although stellar activity may significantly affect estimates of exoplanetary parameters from transit lightcurves \citep{bal12, ber11}, the method used to retrieve the signal of the planet also plays a critical role. The analyses mentioned above were all based on parametric corrections of the instrumental systematics, and are thus, to some degree, subjective.  Recently, non-parametric methods have been proposed to decorrelate the transit signals from the astrophysical and instrumental noise, and ensure a higher degree of objectivity. \cite{wal12, wal13} suggested algorithms based on Independent Component Analysis (ICA) to extract information of an exoplanetary atmosphere from Hubble/NICMOS and Spitzer/IRS spectrophotometric datasets.

In this paper we adopt a similar approach to detrend the transit signal from photometric observations by using Point Spread Functions (PSFs) covering multiple pixels on the detector. We apply this technique to re-analyse the two observations of primary transits of HD189733b recorded with Spitzer/IRAC at 3.6$\mu$m (channel 1 of IRAC) in the ``cold Spitzer'' era. We present a series of tests to assess the robustness of the method and the error bars of the parameters estimated.
Critically, by comparing the results obtained for the two measurements, we discuss the level of repeatibility of transit measurements in the IR, limited by the absolute photometric accuracy of the instrument and possible stellar activity effects. We discuss the reliability of our results for orbital and stellar parameters in the light of previous multiple 8$\mu$m observations \citep{agol10}.

\section{Data analysis}
\label{sec:analysis}

\subsection{Observations}

The two Spitzer observations of HD189733 b discussed here were performed on 2006 October 31 (ID 30590), and 2007 November 25 (ID 40732).

The first observation consists of 1936 exposures using IRAC's stellar mode (full-array), taken over 4.5 hr; 1.8 hr  on the primary transit of the planet, 1.6 hr before, and 1.1 hr after transit. The reset time is 8.4 s. During the observation, the centroid of the star HD189733 was stable to within one pixel.

The second observation was of 1920 exposures using IRAC's sub-array mode, over 4.5 hr; 1.8 hr were spent on the primary transit of the planet, 1.7 hr before, and 1 hr after transit. The interval between consecutive exposures is 8.4 s. Each exposure consists of 64 reads at high speed cadence of 0.1 s. Only for the observation ID 40732, we replaced the 64 reads of each exposure with their mean values, in order to have a manageable number of data points, to reduce the random scatter, and to have the same sampling of the observation ID 30590. During the observation, the centroid of the star HD189733 was again stable to within one pixel. \\

\subsection{Independent Component Analysis in the context of exoplanetary transits lightcurves}
\label{ssec:ica}

Independent Component Analysis (ICA) consists of a transformation from a set of recorded signals to an equivalent set of maximally independent components. The underlying assumptions are that:
\begin{enumerate}
\item each recorded signal is a linear combination of the same source signals;
\item the source signals are mutually independent.
\end{enumerate}
We can express this model as:
\begin{equation}
\label{eqn:ica_system}
\begin{array}{c}
x_1 = a_{1,1} s_1 + a_{1,2} s_2 + ... + a_{1,n} s_n \\
x_2 = a_{2,1} s_1 + a_{2,2} s_2 + ... + a_{2,n} s_n \\
\vdots \\
x_n = a_{n,1} s_1 + a_{n,2} s_2 + ... + a_{n,n} s_n 
\end{array}
\end{equation}
where $x_i$, $i=1 \dots n$, are the recorded signals, $s_j$, $j=1 \dots n$, are the source signals, and $a_{i,j}$ are numerical coefficients.
Eq. \ref{eqn:ica_system} can be written in matrix form as:
\begin{equation}
\label{eqn:ica_direct}
\textbf{x} = \textbf{A} \textbf{s}
\end{equation}
where $\textbf{x}$ is the column vector containing the recorded signals, $\textbf{s}$ is the column vector containing the source signals, and $\textbf{A}$ is the matrix of the coefficients, the so-called ``mixing matrix''.

The aim of ICA is the `blind' separation of the source signals from the observations, i.e., without any additional information (except the assumed mutual independence of the source signals). In other words, the ICA algorithms search for the matrix $\textbf{W}$ that transforms the recorded signals such that the mutual statistical independence is maximised:
\begin{equation}
\textbf{W} \textbf{x} = \textbf{W} \textbf{A} \textbf{s}
\end{equation}
If the assumptions are valid, then $\textbf{W} \textbf{A} = \textbf{D}$, where $\textbf{D}$ is a diagonal matrix, so that:
\begin{equation}
\label{eqn:ica_inversion}
\textbf{W} \textbf{x} = \textbf{D} \textbf{s}
\end{equation}
The diagonal matrix $\textbf{D}$ means that the extracted signals can be rescaled without  changing the mutual independence. \\
To maximise said independence, several approaches and implementations have been proposed \citep{hyv01, tic08}. We used the MULTICOMBI algorithm \citep{tic08}, which optimally mixes EFICA and WASOBI, based on maximising the nongaussianity of the extracted signals \citep{kol06}, and their temporal decorrelations \citep{yer00}, respectively.

In this work, the observed signals are lightcurves of a star, recorded for a time interval that includes an exoplanetary transit event. These lightcurves contain at least three independent contributing signals:
\begin{itemize}
\item the astrophysical signal;
\item the signal of instrumental systematics;
\item stochastic noise.
\end{itemize}
It is possible, in principle, to decompose further the astrophysical and instrumental systematics signals. The former is the sum of the transit signal, the astrophysical background, possible stellar activity signals, etc.; the latter is the sum of different effects from different parts of the instrumentation. All these signals are expected to be independent from each other as they have different origins. By contrast, their linear combinations (i.e. the observed lightcurves) are clearly not mutually independent. It is worth stressing that to disentangle effectively all these signals we need, at least, the number of available lightcurves to be equal to the number of signals. Therefore, we need lightcurves recorded with the same instrument (since lightcurves recorded with different instruments have different systematics plus the astrophysical signals, so that the number of source signals is greater than the number of lightcurves). In principle, using lightcurves recorded at different times with the same instruments should not work, since the systematics have the same origins; but the relevant signals are not necessarily in phase, and so may differ by more than a simple scaling factor. Additionally, further differences might be present due to stellar variability. However, the transit signal, being common to all the lightcurves, is potentially detrendable. A successful extraction of a transit signal from a time series spanning several exoplanetary transit events, conveniently split into sub-lightcurves, is described in \cite{wal12}.\\
The advantage of spectroscopic observations over photometry is the provision of simultaneous lightcurves at different wavelengths with largely common instrumental systematics. The transit signals at each wavelength can be obtained by subtracting proper systematics models from the lightcurves (an accurate direct extraction of the transit is impossible due to the limb darkening effect). By using this technique, \cite{wal13} have extracted an infrared transmission spectrum of HD189733b between 1.51 $\mu$m and 2.43 $\mu$m, from a Hubble/NICMOS dataset.

\subsection{ICA using pixel-lightcurves}
\label{ssec:the_method}

The main novelty of the algorithms we use here is their ability to detrend the transit signal from a single photometric observation of just one primary transit. This is possible because, even if stars can be approximated by point sources, the instrument is purposely de-focused to spread the PSF over several detector pixels, and the position of the target star on the detectors is stable to within one pixel. During an observation, there are several pixels detecting the same astrophysical signals at any time, but with different scaling factors, depending on their received flux, their quantum efficiency, and the instrument PSF. \\
We performed an ICA decomposition over several pixel-lightcurves, i.e. the time series from individual pixels, in order to extract the transit signal and other independent signal components (stellar or instrumental in nature). \\
Once a set of independent components has been obtained from a selected set of pixel-lightcurves, different approaches to obtain the transit signal can be considered.

\textit{\underline{Method 1}: direct identification of the transit component} \\
In principle, if one of the independent components extracted has the morphology of the transit signal, we assume that one to be the transit signal, multiplied by an undetermined scaling factor. We renormalise the signal by the mean value calculated on the out-of-transit part, so that the out-of-transit level is unity. \\
Method 1 is not applicable to the extraction of accurate transit signals from spectroscopically resolved observations of a primary transit at different wavelengths, because of the wavelength dependence of stellar limb darkening. This is not a problem in our case, because all the pixels record the same wavelengths. \\

\textit{\underline{Method 2}: non-transit-components subtraction} \\
Another approach to estimating the transit signal is to remove all the other effects from an observed lightcurve, i.e. by subtracting all the components other than the transit one, properly scaled. The scaling factors can be determined by fitting a linear combination of the components, plus a constant term, to the out-of-transit part of the lightcurve \footnote{The out-of-transit limits do not have to be known exactly. They can be chosen in a way to be sure of not including part of the transit while fitting, relying on parameters reported in previous papers and on the lightcurves themselves. The results should not be affected by this choice, but it is worth checking this point.}. The coefficients of the linear combination and the constant are the free parameters to fit. \\
Instead of fitting the non-transit-components on the pixel-lightcurves, and then subtracting, we performed these processes on the spatially integrated lightcurves, obtained by summing all the individual pixel-lightcurves. The integrated lightcurves are much less noisy than the individual pixel-curves. \\

\subsection{Transit lightcurve fitting and error bars}
\label{ssec:curvefit}


After the extractions of the detrended and normalised transit time series, we modelled them by using the \cite{ma02} analytical formulae. We can compute the observed flux as a function $F(p,z)$, where $p=r_p/R_s$ is the ratio between the planetary and the stellar radii, and $z=d/R_s$ is the distance between the centres of their disks projected onto the sky divided by the stellar radius.
The relative distance $z$ is a function of time, determined by the orbital parameters.\\
We assumed the orbital period $P$, zero eccentricity, and a quadratic limb darkening model \citep{how11}.
The values of the fixed parameters are reported in Tab. \ref{tab1}.
\begin{table}[!h]
\begin{center}
\caption{Values of the parameters fixed while generating the transit models. The limb darkening coefficients, $\gamma_1$ and $\gamma_2$, were computed for a star with effective temperature $T_{eff} = 5000 K$, gravity $\log{g} = 4.5$, mixing-length parameter $l/h = 1.25$, solar abundances. \label{tab1}}
\begin{tabular}{cc}
\tableline\tableline
$P$ & $2.218573 \ days$\\
$e$ & $0$\\
$\gamma_1$ & $7.82118 \times 10^{-2}$\\
$\gamma_2$ & $2.00656 \times 10^{-1}$\\
\tableline
\end{tabular}
\end{center}
\end{table}
\\
We first determined the centers of the transit ephemeris by fitting some symmetric models with all the other parameters fixed. Recent papers \citep{col10, tri10} report a small but non-zero eccentricity ($e \simeq 4 \cdot 10^{-3}$), but we verified this would affect our estimates of the other parameters by a negligible fraction of their error bars. \\
We then performed a fit with three free parameters:
\begin{enumerate}
\item the ratio of planetary to stellar radii, $p = r_p/R_s$;
\item the orbital semimajor axis (in units of the stellar radius), $a_0 = a/R_s$;
\item the orbital inclination, $i$.
\end{enumerate}
We chose these as free parameters, because:
\begin{itemize}
\item there is a large range of values published in the literature;
\item they largely determine the shape of the transit signal;
\item they do not show strong cross-correlations.
\end{itemize}
For completeness, and for comparisons with the literature, in the final results we report also the transit depth, $p^2$, the impact parameter, $b$, and the duration of the transit, $T$, where
\begin{equation}
b = a_0 \cos{i}
\end{equation}
\begin{equation}
T = \frac{P \sqrt{1-b^2}}{ \pi a_0}
\end{equation}

We used a Nelder-Mead optimisation algorithm \citep{lag98}, to obtain first estimates of the parameters of a model. To confirm/improve these estimates and to determine error bars, we ran an Adaptive Metropolis algorithm with delayed rejection \citep{haa06} for 20,000 iterations, starting from the optimal values initially determined, in order to sample the probability distributions of the fitted parameters. The updated best estimates and error bars of the parameters are the means and the standard deviations of the sampled distributions (approximately gaussians), respectively. No burn-in is required, because of the optimal starting points of the chains. \\
The variance of the likelihood function is initialised as the variance of the residuals obtained for the first model and then sampled together with the other free parameters ($\sigma_0^2$).
In this way, we take into account both white and the autocorrelated noise present in the detrended time series, but we ignore possible systematic errors due to the preliminary ICA deconvolution. The ICA errors can be represented as an additional uncertainty, $\sigma_{ICA}$, on each point in the time series. The likelihood's variance, $\sigma_{like}^2$, becomes:
\begin{equation}
\sigma_{like}^2 = \sigma_0^2 + \sigma_{ICA}^2
\end{equation}
We tested that resampling the parameters' chains with $\sigma_{like}^2$ does not affect their best values, while the total error bars of the single parameters, $\sigma_{par}$, increase with respect to the previous estimates (without the ICA errors), $\sigma_{par,0}$, as:
\begin{equation}
\sigma_{par} = \sigma_{par,0} \frac{ \sigma_{like}}{ \sigma_{0}} = \sigma_{par,0} \sqrt{ \frac{ \sigma_{0}^2 + \sigma_{ICA}^2 }{ \sigma_{0}^2}}
\end{equation}
A measure of the uncertainties on the independent components extracted by ICA is given by the Interference-to-Signal-Ratio matrix, $\textbf{ISR}$, i.e. a $n \times n$ matrix, where $n$ is the number of signals. The $\textbf{ISR}_{ij}$ element estimates the relative remaining presence of the $j^{th}$ component in the $i^{th}$ one. Then,
\begin{equation}
\textbf{ISR}_i = \sum_{j=1, \ j \ne i}^{n} \textbf{ISR}_{ij}
\end{equation}
estimates the relative remaining presence of all the other components in the $i^{th}$ one. \\
If the $i^{th}$ component represents the transit signal, and if we estimate the transit signal through \textit{method 1}, we can identify:
\begin{equation}
\sigma_{ICA}^2 = f^2 \textbf{ISR}_i
\end{equation}
$f$ being the scaling factor used. \\
If the $i^{th}$ component represents the transit signal, but we estimate it through \textit{method 2}, $\sigma_{ICA}$ has to contain a weighted sum of the \textbf{ISR}s of the non-transit components removed, plus the discrepancies of the fit to the out-of-transit phases:
\begin{equation}
\label{eqn:sigmaica2}
\sigma_{ICA}^2 = f^2 \left ( \sum_{j=1}^{m} o_j^2 \textbf{ISR}_j + \sigma_{ntc-fit}^2 \right )
\end{equation}
$o_j$ being the coefficients of the non-transit components, $m$ the number of components considered, $\sigma_{ntc-fit}$ the standard deviation of the residuals from the reference lightcurve (out of the transit), and $f$ the normalising factor for the model-subtracted lightcurve. \\
The MULTICOMBI code produces two Interference-to-Signal-Ratio matrices, $\textbf{ISR}^{EF}$, associated with the algorithm EFICA, and the $\textbf{ISR}^{WA}$, associated with the algorithm WASOBI. We estimated the global $\textbf{ISR}$ as their average:
\begin{equation}
\textbf{ISR} = \frac{ \textbf{ISR}^{EF} + \textbf{ISR}^{WA}}{2}
\end{equation}
This is a conservative estimate, given that, according to \cite{tic08}, the MULTICOMBI $\textbf{ISR}$ slightly outperforms the best of $\textbf{ISR}^{EF}$ and $\textbf{ISR}^{WA}$ (then it could be smaller), but these estimates are entirely reliable only under certain assumptions on the signals which may be not satisfied in these cases. Here we take them as worst-case estimates.

\subsection{Application to observations}
\label{ssec:applyID30590}

Here, we describe the main steps of the analyses performed on each of the two observations (ID 30590 and ID 40732), which include some tests of robustness. We now discuss only results obtained with method 2, as they are much more stable; results obtained with method 1 are reported in Appendix \ref{sec:app2}, along with a critical comparison of the two methods.

\subsubsection{Choice of the pixels}
\label{ssec:pixel_choice}

The first step in the analysis is the choice of the pixel-lightcurves to analyse. This is determined by:
\begin{itemize}
\item the instrument point response function (PRF), i.e. the measured intensity profile of the star on the detector  \footnote{Note that the PRF is, in principle, slightly different to the PSF: the PSF is the intensity profile incident on the detector, while the PRF is the measured intensity profile (including the detector response).};
\item the noise level of the detector;
\item the effective number of significant components to disentangle.
\end{itemize}
The number of significant components is not known a priori. The ICA code extracts a number of components equal to the number of lightcurves that it receives as input. Apart from the collective behaviour common to all the pixel-lightcurves, each pixel introduces an individual signature. Only if the individual signatures are negligible compared to the collective behaviour are we able to select enough lightcurves to disentangle the significant components. The PRF and the noise level of the detector limit the number of pixels containing potentially useful astrophysical information.

\begin{figure}[!b]
\epsscale{.80}
\plotone{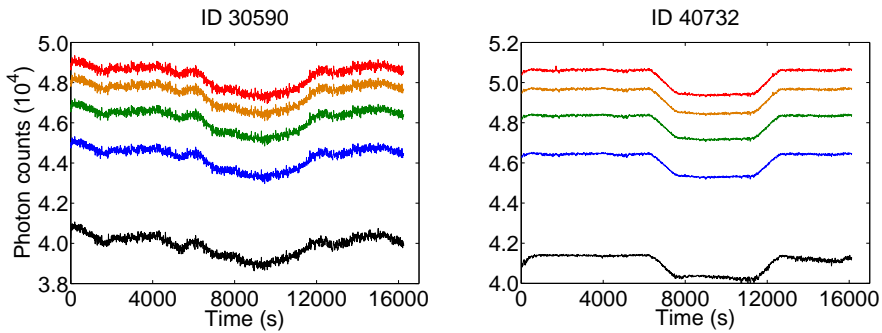}
\caption{Raw integral lightcurves from several squared arrays of pixels: black $3 \times 3$, blue $5 \times 5$, green $7 \times 7$, orange $9 \times 9$, and red $11 \times 11$ (in order of increasing counts). \label{fig1}}
\end{figure}

In practice, we considered several arrays of pixels with the stellar centroid at their centers, having dimensions $3 \times 3$, $5 \times 5$, $7 \times 7$, $9 \times 9$, and $11 \times 11$ pixels. Fig. \ref{fig1} shows the ``integral lightcurves'', obtained by summing the contributions from the various pixels. We looked for outliers in the time series, i.e. points discrepant more than 5$\sigma$ from a first transit-lightcurve model (fitted on the original data), and we replaced those outliers with the averages of the points immediately before and after. We found  only one outlier in observation ID 30590, and nine in ID 40732. Although the observed lightcurves are two primary transits of the same exoplanet, observed at the same wavelength through the same instrument, they appear very different, mainly because of the different observing strategies. In particular, observation ID 40732 seems to be much less affected by systematics, and less noisy.
The integral lightcurves from the various arrays of pixels look very similar in shape, but have different absolute intensities, as expected. The mean intensities of the integral $3 \times 3$, $5 \times 5$, $7 \times 7$ and $9 \times 9$ lightcurves are respectively $\sim 83 \%$, $\sim 92 \%$, $\sim 96 \%$ and $\sim 98 \%$ of the mean intensity of the integral $11 \times 11$ lightcurve. We are not interested in absolute photometry, but only in relative variations of the intensity, therefore it is not important whether the PRF is totally contained in the square used for the analysis or not, provided it contains enough information to detrend the transit signal. Larger arrays include pixels which add noise with little or no astrophysical information. We concluded that the $3 \times 3$ or the $5 \times 5$ arrays were the optimal choices. However, we tested all the pixel arrays, to assess the robustness of the results.

We binned the transit time series by replacing groups of nine consecutive points with their mean values, in order to reduce the computational time required to sample the parameters' distributions in the \cite{ma02} model (see Sec. \ref{ssec:curvefit}). We checked that in select cases this approach does not affect the parameter estimates.

The best values of $p$, $a_0$, and $i$ are stable, within the error bars, with respect to the choice of the set of pixel-lightcurves used to detrend the signals. The discrepancies between the extracted signals and the relative fits are the biggest for the $3 \times 3$ array; for larger arrays they are smaller, and are either all at the same level (ID 30590), or slightly decrease with the size of the array (ID 40732). Our interpretation of this is that the $5 \times 5$ and larger arrays contain the same amount of useful information, while in the $3 \times 3$ array something is missed.  The ICA errors confirm this hypothesis, being the smallest for the $7 \times 7$ (ID 30590) and $5 \times 5$ (ID 40732) arrays. Higher values for larger arrays were expected, but do not differ significantly. We conclude that the choice of the array size is not crucial.

\subsubsection{Choice of the components}
\label{ssec:components_ID30590}

\begin{figure}[!b]
\epsscale{.80}
\plotone{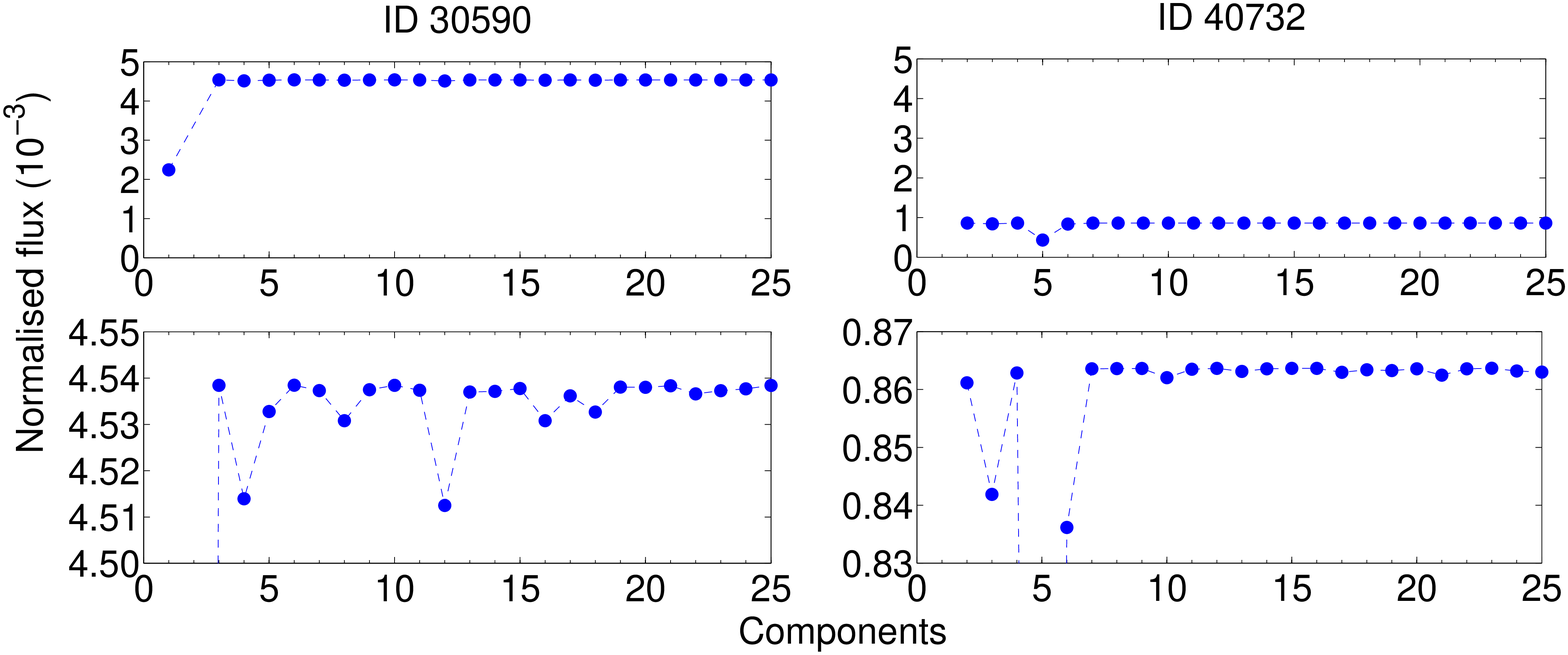}
\caption{Top: Standard deviations of the residuals of the single-component fits, normalised to out-of-transit level; $5 \times 5$ array. Bottom: the same, zooming on the topmost part of the curve.
\label{fig2}}
\end{figure}

\begin{figure}[!b]
\epsscale{.80}
\plotone{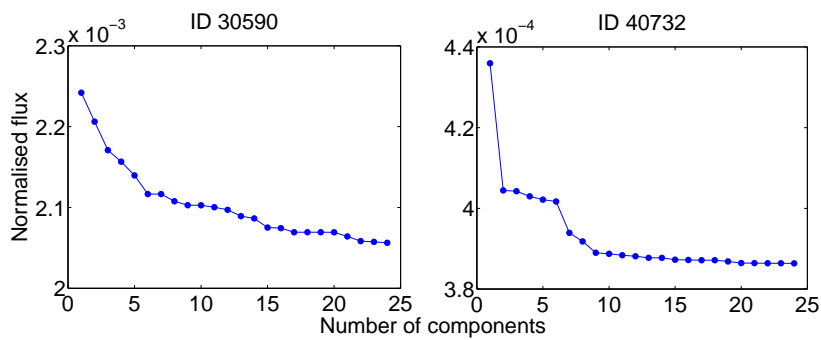}
\caption{Standard deviations of the residuals of the fits with multiple components, normalised to out-of-transit level, $5 \times 5$ array. \label{fig3}}
\end{figure}

In Sec. \ref{ssec:pixel_choice}, we corrected the observed lightcurves by subtracting all the non-transit components (see Sec. \ref{ssec:the_method}). Here, we show how to identify the most significant components, and how many should be considered. We generally expect that some components are related to collective behaviours, common to all the pixels, and others to individual pixels' signatures and/or noisy mixtures of the sources. By inspection, a few of the components clearly present time structures, while others are random scattered time series. \\
We report results from the $5 \times 5$ array only, as it is the smallest array containing all the astrophysical and instrumental information. \\
To evaluate the impact of each component in the out-of-transit data, we found the best fits of the single components (plus additive constants) to that part of the integral lightcurve, and calculated the means and standard deviations of the residuals. In this way, we established a ranking of importance of the components, based on the minimisation of the discrepancies between their fits and the integral lightcurve, out of transit. We then computed other best fits by using the $n$ most important components, according to that ranking, with $n$ from $1$ to $24$. The best coefficients for the components and the additive constants were determined through the Nelder-Mead optimisation algorithm. \\
Fig. \ref{fig2} reports the standard deviations of the residuals of the single-component fits, normalised to the out-of-transit level; Fig. \ref{fig3} reports analogous fits obtained using more components. Note that figures related to different observations are not reported with the same scale, because they have very different accuracies.
Most systematics are contained in one major component, but the use of multiple components increases the detrending accuracy.

\begin{figure}[!h]
\epsscale{.80}
\plotone{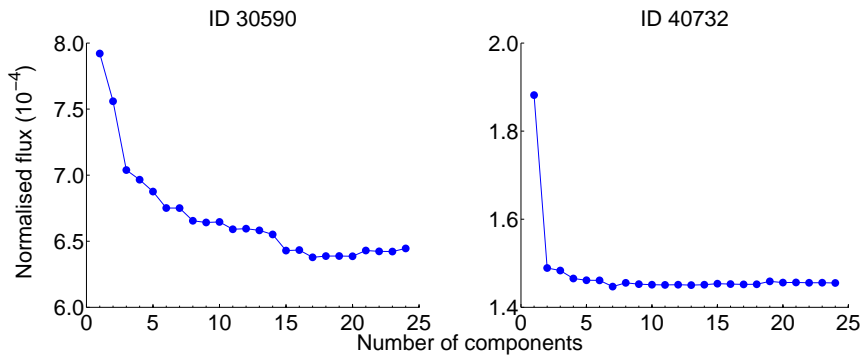}
\caption{Standard deviations of the residuals between the transit signals estimated using method 2 (with the $n$ most important components, binned by nine points), and the corresponding best model fits. \label{fig4}}
\end{figure}

\begin{figure}[!h]
\epsscale{.80}
\plotone{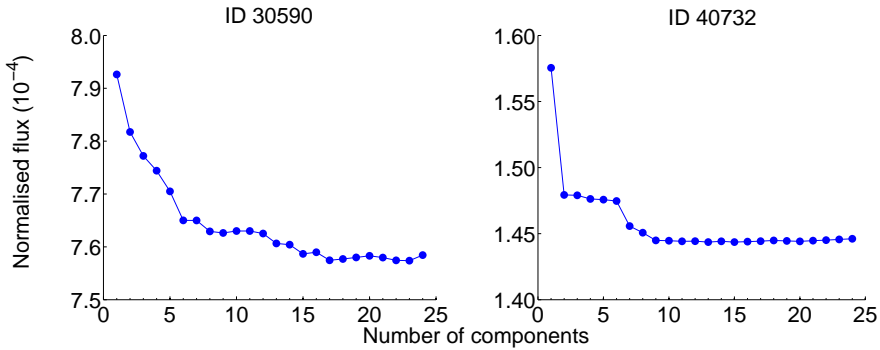}
\caption{ICA separation errors for the transit signals estimated through method 2 (with the $n$ most important components, binned by nine points). \label{fig5}}
\end{figure}

We computed twenty-four estimates of the transit signal by removing the $n$ most significant non-transit components from the integral lightcurve. We binned these over nine points, as in Sec. \ref{ssec:pixel_choice}, and fitted \cite{ma02} models to these curves. The standard deviations of the residuals between each curve and the corresponding best model of \cite{ma02} are reported in Fig. \ref{fig4}, and confirm that the use of multiple components for detrending improves the results. \\
ICA separation errors are plotted in Fig. \ref{fig5}, showing the same trends.

Given these tests, we expect to have good estimates of the transit signals by removing the first few most significant components, but some improvements can be made by removing more components, up to a saturation point.
The best values of the parameters $p$, $a_0$, and $i$, for each estimated transit signal, are shown in Fig. \ref{fig12}.

The dispersions in the parameters are fully contained in the intervals previously estimated by using the signals with all non-transit components removed (see Appendix \ref{sec:app0}, Tab. \ref{tab6}, column 2), except for values from the transit signal from observation ID 40732 with only one component removed.

\section{Results}

\begin{figure}[!h]
\epsscale{0.80}
\plotone{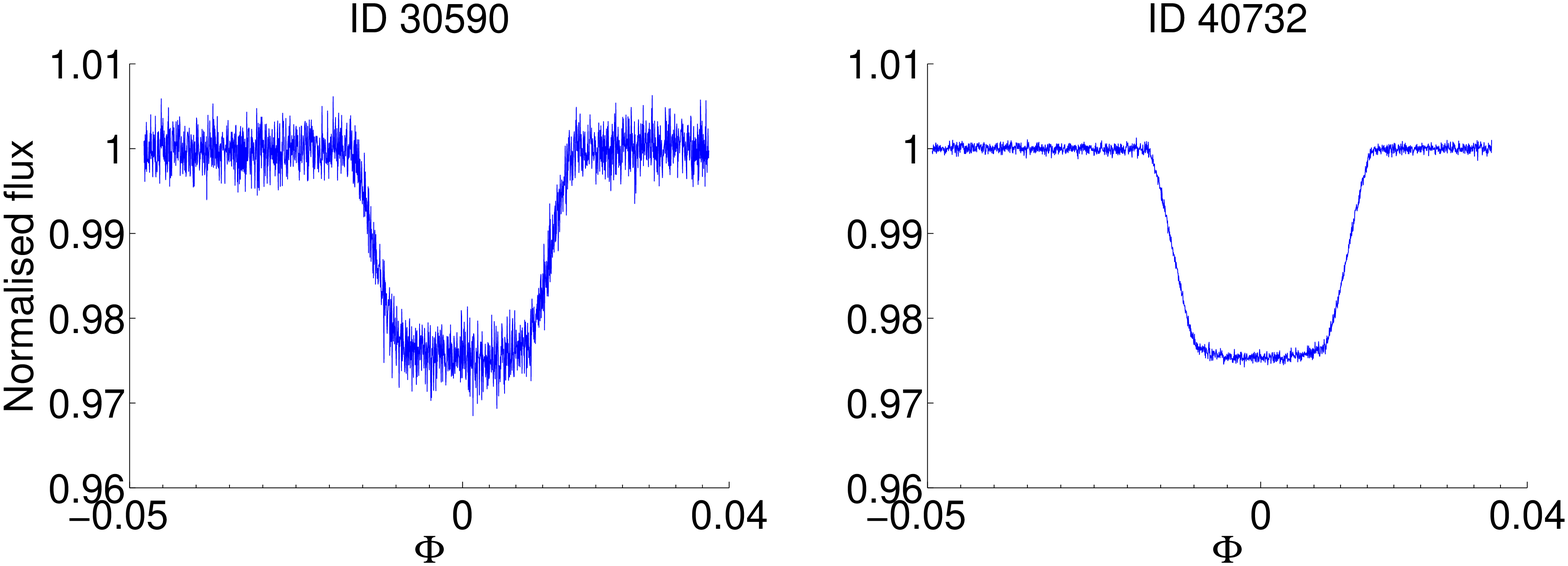}
\plotone{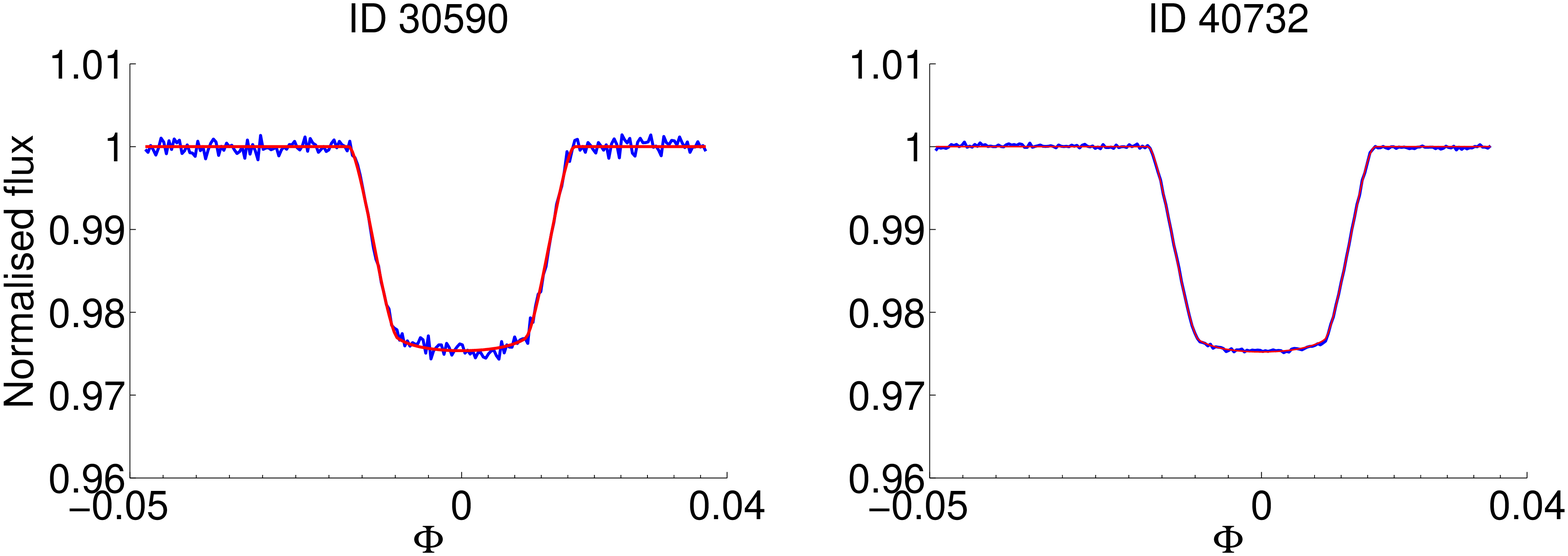}
\plotone{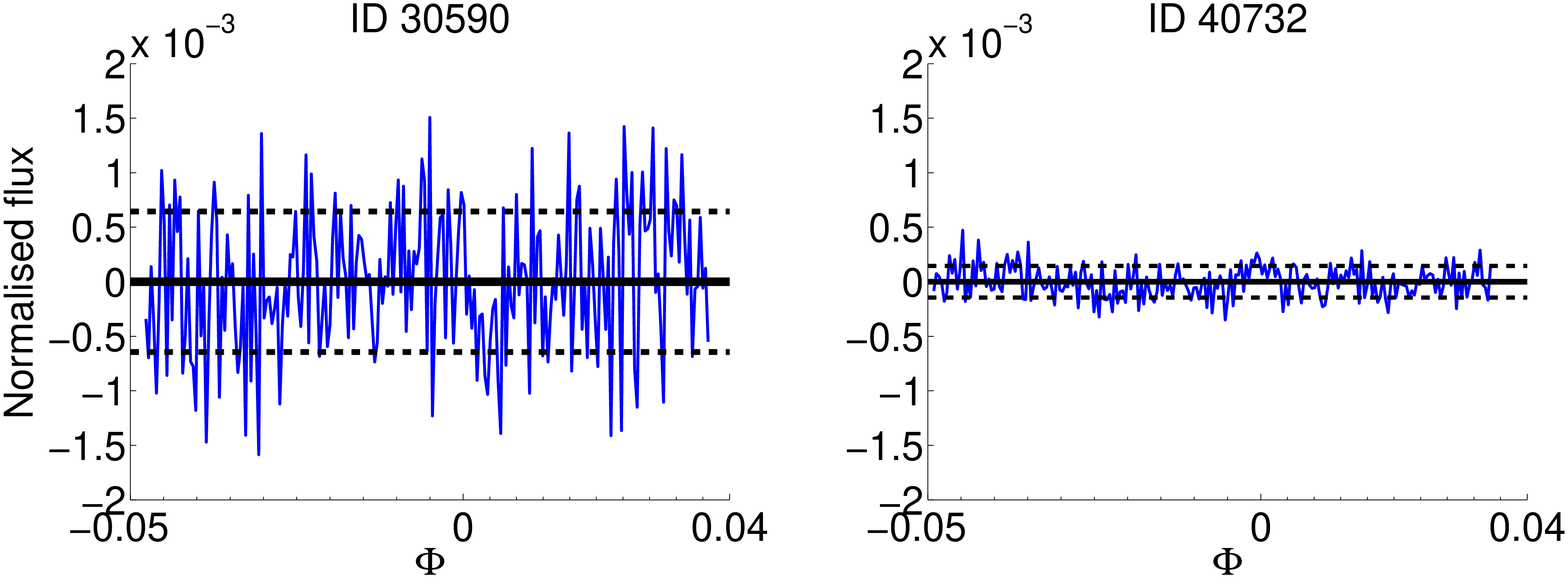}
\caption{(Top panel): transit time series extracted using the $5 \times 5$ array, considering all the independent components (see Sec. \ref{ssec:the_method}). (Middle panel): (blue) the same series, binned by nine points, (red) relative best model fit. (Bottom panel): residuals between the extracted time series and the model. Dashed black lines indicate the standard deviations of the residuals. \label{fig6}}
\end{figure}

Fig. \ref{fig6} shows the normalised transit signals extracted using the $5 \times 5$ arrays, considering all the independent components; the relative best lightcurve fits to the binned and detrended data; and the residuals. The standard deviations of the residuals are $\sigma_0^{ID 30590} = 6.4 \times 10^{-4}$ and $\sigma_0^{ID 40732} =1.45 \times 10^{-4}$. Note that the signal extracted from observation ID 40732 has a dispersion smaller by a factor $\sim 4.4$.\\

\begin{figure}[!h]
\epsscale{.80}
\plotone{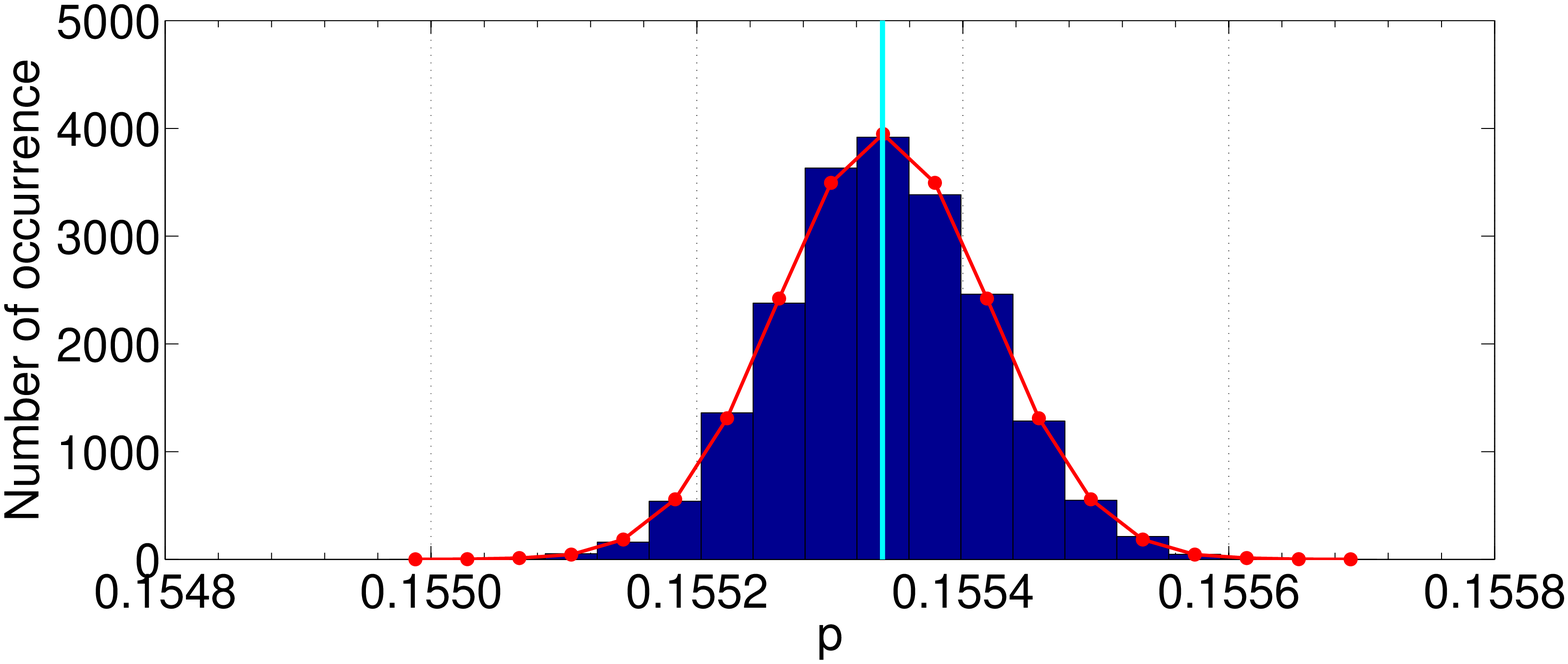}
\plotone{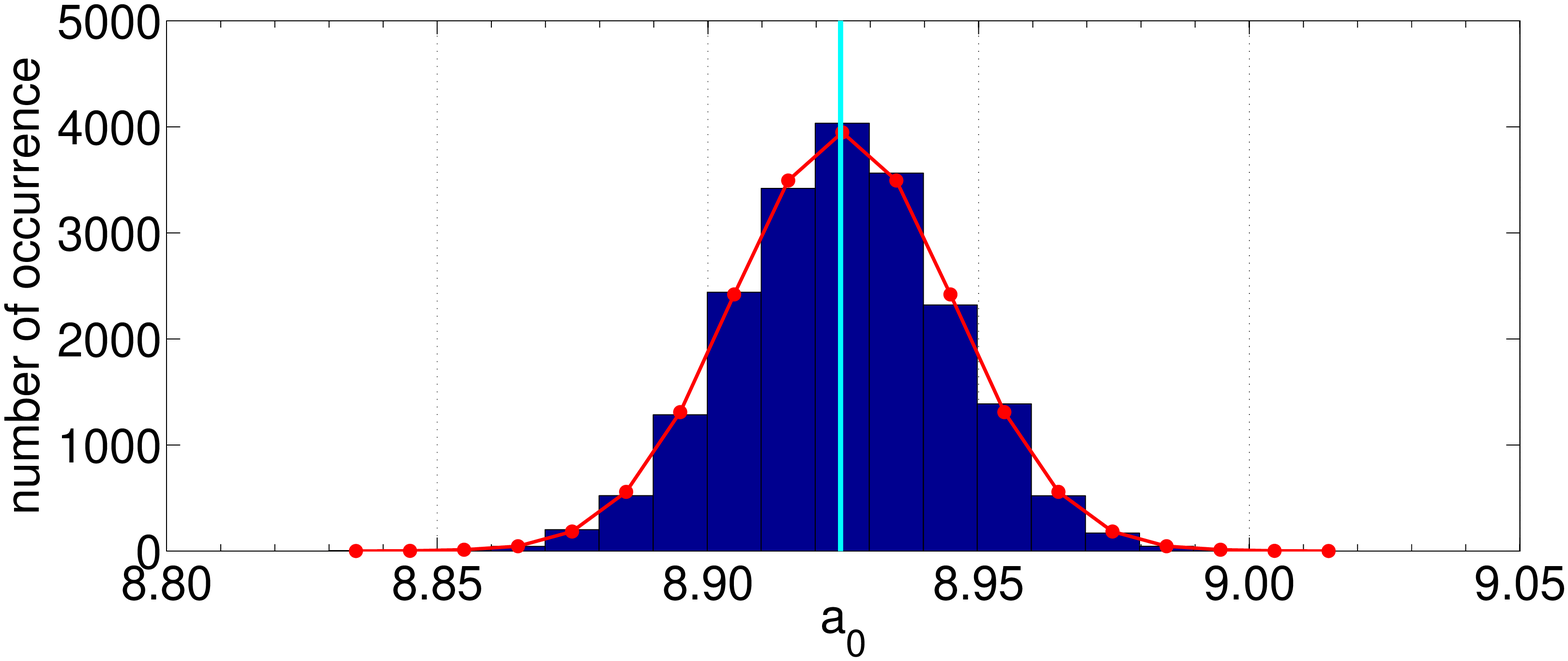}
\plotone{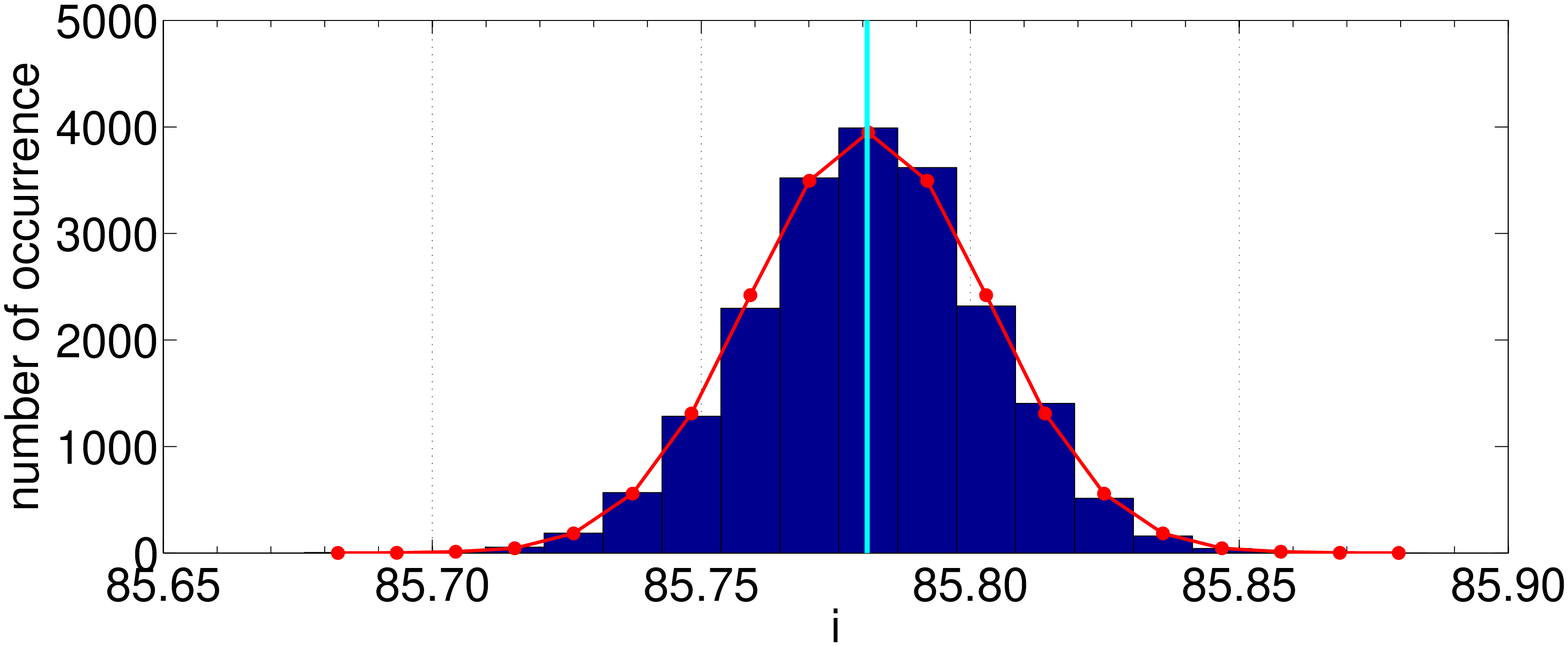}
\caption{From top to bottom, histograms of the sampled chains for parameter $p$, $a_0$, and $i$, respectively, relative to the time series estimated from the $5 \times 5$ array, considering all the independent components, without including the ICA error (observation ID 40732). The overplotted red curves show gaussian distributions with the sampled means and variances. The light-blue vertical lines indicate the starting values, determined through the Nelder-Mead optimisation (see Sec. \ref{ssec:the_method} and \ref{ssec:curvefit}). \label{fig7}}
\end{figure}

Fig. \ref{fig7} illustrates the sampled distributions of the parameters $p$, $a_0$, and $i$, from the transit signal extracted by observation ID 40732 (see Sec. \ref{ssec:curvefit}). Similar distributions, but with larger dispersions, were obtained for the other transit signal.
Tab. \ref{tab2} reports the starting values, sampled means, and standard deviations of the parameters obtained.
\begin{table}[!h]
\begin{center}
\caption{Estimated $p$, $a_0$, and $i$ from the $5 \times 5$ array of of pixels, through method 2, considering all the independent components, without including the ICA error (see Sec. \ref{ssec:the_method} and \ref{ssec:curvefit}). \label{tab2}}
\begin{tabular}{cccc}
\tableline\tableline
 & $Starting \ value$ & $Mean$ & $Standard \ deviation$\\
\tableline
ID 30590 & & & \\
$p$ & $0.15471$ & $0.15470$ & $3.6 \cdot 10^{-4}$\\
$a_0$ & $9.05$ & $9.06$ & $0.09$\\
$i$ & $85.93$ & $85.94$ & $0.10$\\
\tableline
ID 40732 & & & \\
$p$ & $0.15534$ & $0.15534$ & $8 \cdot 10^{-5}$\\
$a_0$ & $8.92$ & $8.92$ & $0.02$\\
$i$ & $85.78$ & $85.78$ & $0.02$\\
\tableline
\end{tabular}
\end{center}
\end{table}
Note that the starting values agree very well with the sampled means.
The likelihood variances without the ICA contribute, calculated as detailed in Sec. \ref{ssec:curvefit}, are equal to the variances of the residuals: $\sigma_{0}^{ID 30590} = \left ( 6.5 \pm 0.3 \right ) \times 10^{-4}$, and $\sigma_{0}^{ID 40732} = \left ( 1.46 \pm 0.07 \right ) \times 10^{-4}$.

Tab. \ref{tab3} gives the final results for the parameters $p$, $a_0$, $i$, $p^2$, $b$, and $T$. \\
\begin{table}[!h]
\begin{center}
\caption{Adopted best estimates of parameter values. \label{tab3}}
\begin{tabular}{ccc}
\tableline\tableline
 & ID 30590 & ID 40732\\
\tableline
$p$ & $0.1547 \pm 0.0005$ & $0.15534 \pm 0.00011$\\
$a_0$ & $9.05 \pm 0.16$ & $8.92 \pm 0.03$\\
$i$ & $85.93 \pm 0.15$ & $85.78 \pm 0.03$\\
$p^2$ & $0.02394 \pm 0.00017$ & $0.02413 \pm 0.00003$\\
$b$ & $0.64 \pm 0.03$ & $0.657 \pm 0.005$\\
$T$ & $5170 \pm 200 \ s$ & $5157 \pm 34 \ s$\\
\tableline
\end{tabular}
\end{center}
\end{table}

\subsection{Combining observations}
\label{ssec:combo}

\begin{figure}
\epsscale{.80}
\plotone{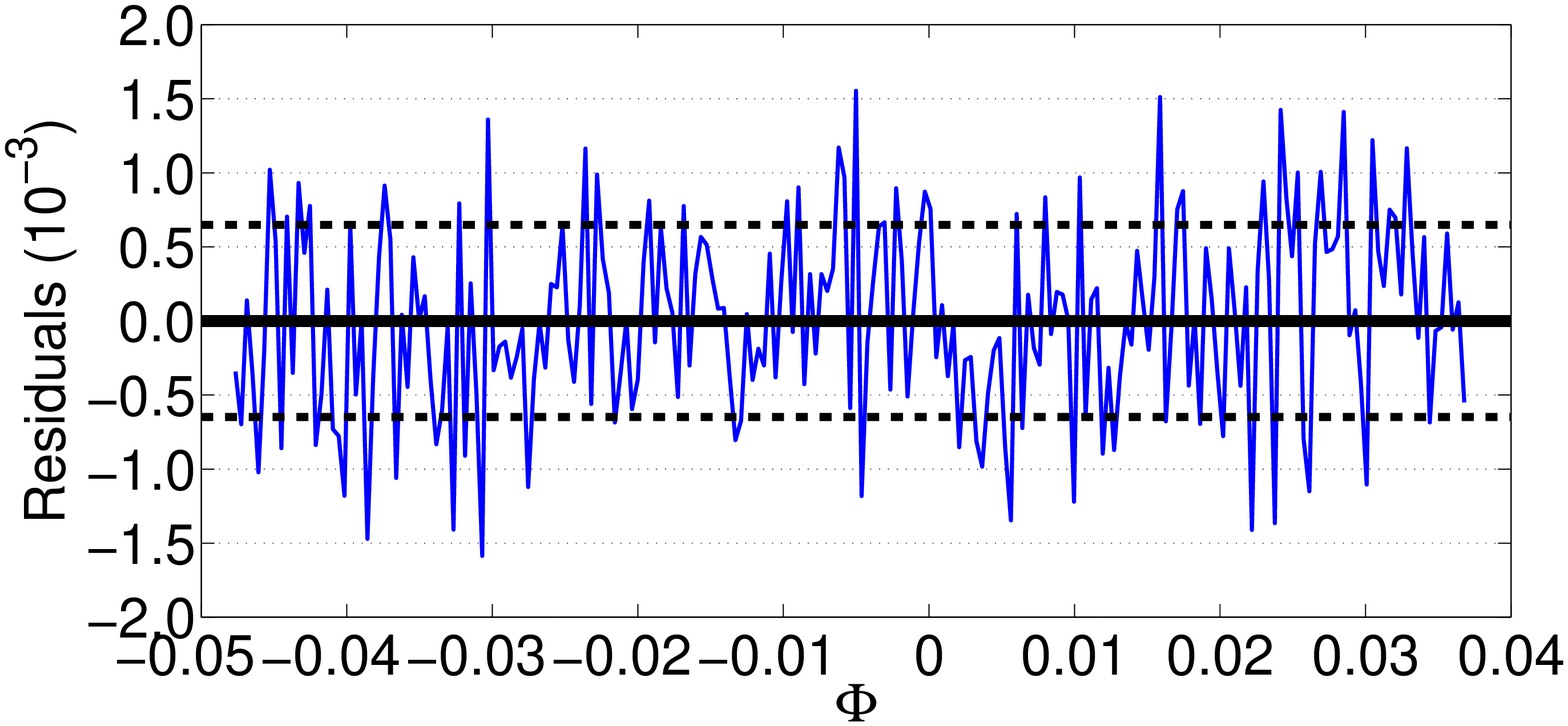}
\caption{Residuals between the transit signal from observation ID 30590 and the model fit with $a_0 = 8.92$ and $i = 85.78$. Black dashed lines indicate the standard deviation, which is consistent with the standard deviation of the residuals between the signal and the model estimated with $a_0$ and $i$ as free parameters. \label{fig8}}
\end{figure}

\begin{figure}
\epsscale{.80}
\plotone{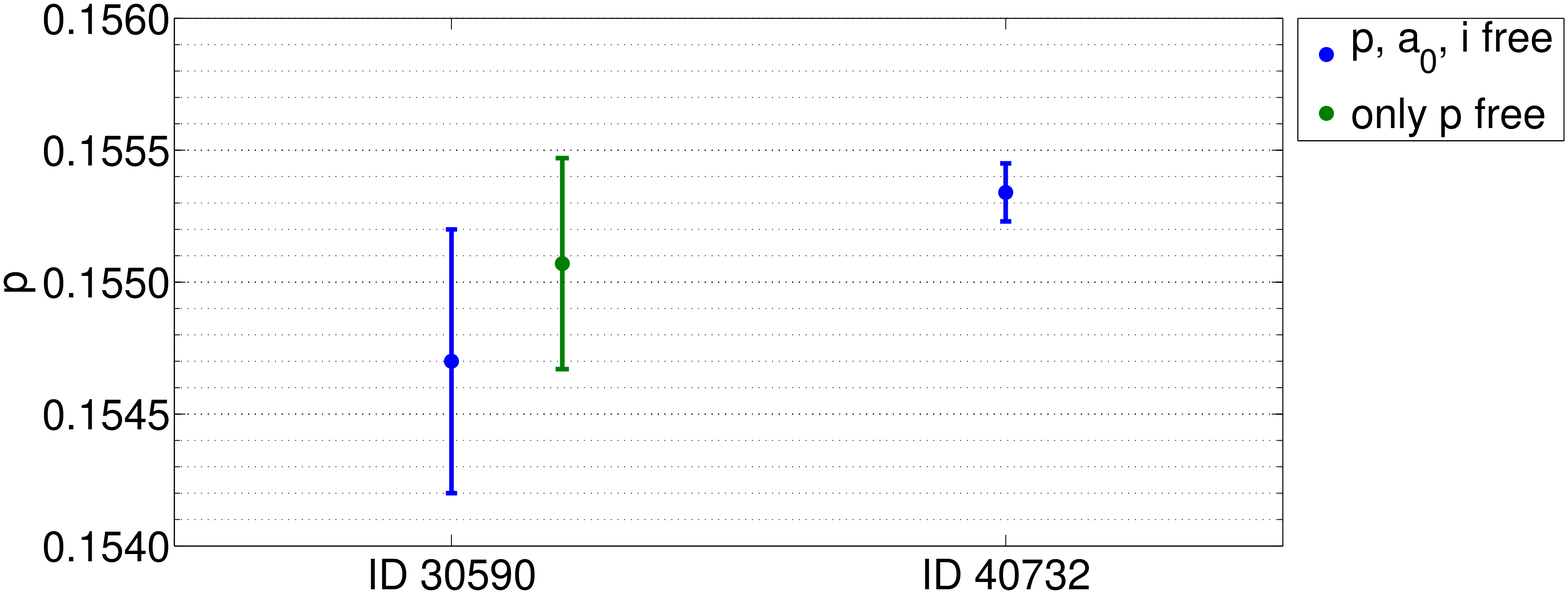}
\plotone{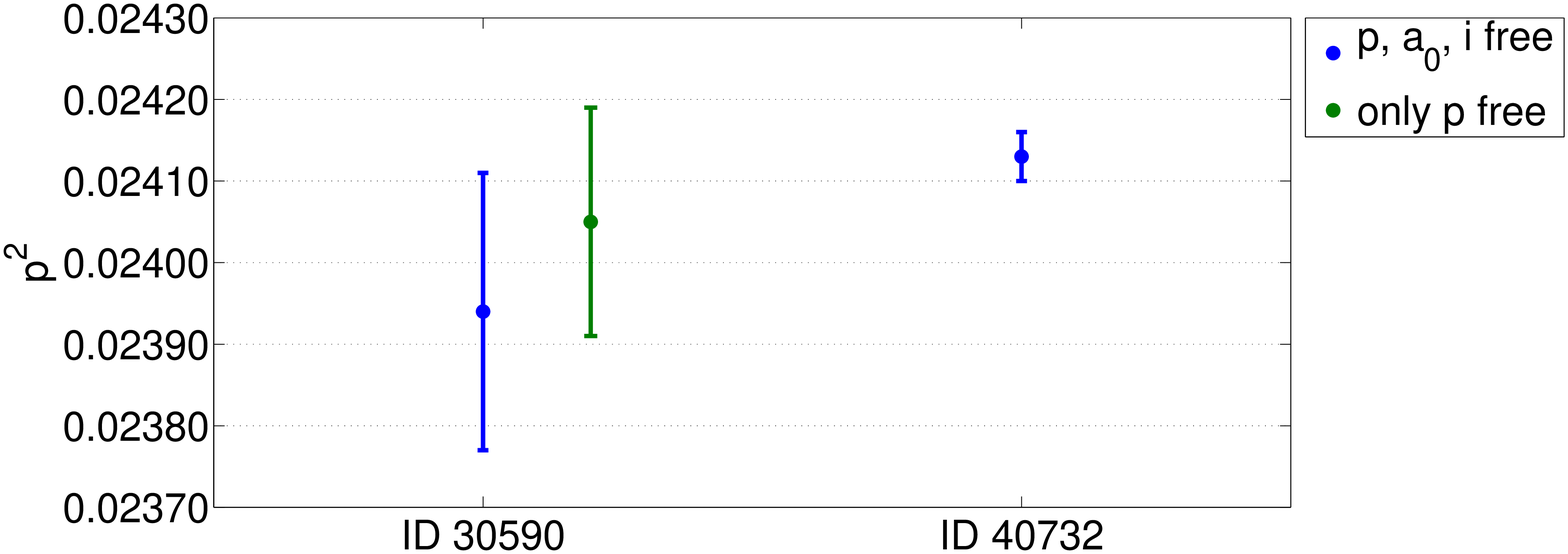}
\caption{Top: Estimates of $p$; with $a_0$ and $i$ free (blue); with $a_0 = 8.92$, and $i = 85.78$, i.e. the best values found for observation ID 40732 (green). Bottom: the same for $p^2$. \label{fig9}}
\end{figure}

The parameter estimates determined from observation ID 40732 are much more accurate than those from ID 30590. Assuming that the orbital parameters were the same along the two observations, as expected because of the stability of the planetary orbit, we computed a chain for ID 30590, for $p$ only, with $a_0$ and $i$ fixed to the best values estimated from ID 40732. In this way, we can make a direct comparison of $p$ between the two observations, and avoid possible correlations with the other parameters. Even if $a_0$ and $i$ were badly determined, due to an inaccurate stellar model being assumed (e.g. wrong limb darkening coefficients, star spots, or faculae), they would introduce a systematic error on $p$ that would be equal for both observations. Thus variations of $p$ (or $p^2$), obtained while keeping all other parameters fixed, are a more objective measurement of the stellar variations.  Results are reported in Tab. \ref{tab4}; note that $\sigma_{0}$ is unchanged. Fig. \ref{fig8} shows the discrepancies between the detrended signal and the model.
\begin{table}
\begin{center}
\caption{Estimated best values and standard deviations of $p$, from observation ID 30590, with $a_0 = 8.92$, and $i = 85.78$, without including the ICA error. Discrepancies between the signals and the related best model fits (see Sec. \ref{ssec:curvefit}). \label{tab4}}
\begin{tabular}{cc}
\tableline\tableline
$p$ & $0.15507$\\
$\sigma_p^0$ & $2.7 \cdot 10^{-4}$\\
\tableline
$\sigma_{0}$ & $6.5 \cdot 10^{-4}$\\
\tableline
\end{tabular}
\end{center}
\end{table}
Including the ICA errors we found:
\begin{equation}
\begin{array}{c}
p = 0.1551 \pm 0.0004 \\
p^2 = 0.02405 \pm 0.00014
\end{array}
\end{equation}
Fig. \ref{fig9} compares the original estimates for $p$ and $p^2$, with those obtained by keeping $a_0$ and $i$ fixed.
We note that:
\begin{itemize}
\item the best value from ID 30590 with $a_0$ and $i$ fixed agrees better with result from ID 40732;
\item the new estimate from ID 30590 is consistent with the previous one, but with a (slightly) smaller error bar.
\end{itemize}

\section{Discussion}

\subsection{Comparison of the two observations}

\begin{figure}
\epsscale{.80}
\plotone{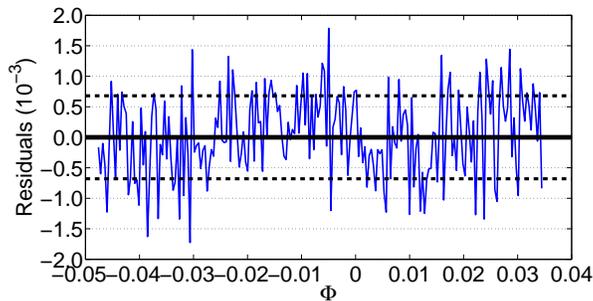}
\caption{Residuals obtained by subtracting observation ID 40732 to observation ID 30590. Black dashed lines indicates their standard deviation. \label{fig10}}
\end{figure}

The planetary, orbital, and stellar parameters derived separately from the two observations are all consistent within 1$\sigma$. In particular, the duration of the transit is extremely stable between the two observations. This is not surprising, because its measure is almost insensitive to calibration errors and stellar activity; all the other parameters are much more affected by these sources of noise. Furthermore, these other parameters are strongly correlated; e.g. a non-optimal estimate of the impact parameter $b$ will result in an imprecise transit depth $p$, etc. Fig. \ref{fig10} shows the differences between the transit signals extracted for the two observations.
The standard deviation of the differences is $\sim 6.8 \times 10^{-4}$, which is comparable with the standard deviation of the discrepancies between the signal from observation ID 30590 and the relative model fit ($\sigma_{0}^{ID 30590} = \left ( 6.5 \pm 0.3 \right ) \times 10^{-4}$); the discrepancies between the signal from observation ID 40732 and the corresponding model fit are negligible. Hence there is no evidence of physical variations in the transit signal from one observation to the other one. \\
The results of Sec. \ref{ssec:combo} reinforce our claim of non detectable stellar activity variations. \\
We conclude that the two observations lead to consistent results, but the second constrains the orbital and stellar parameters much better, and allows the estimate of the transit depth for the first one to be refined.

\subsection{Comparison with observations at $8 \mu m$}

\begin{figure}
\epsscale{.80}
\plotone{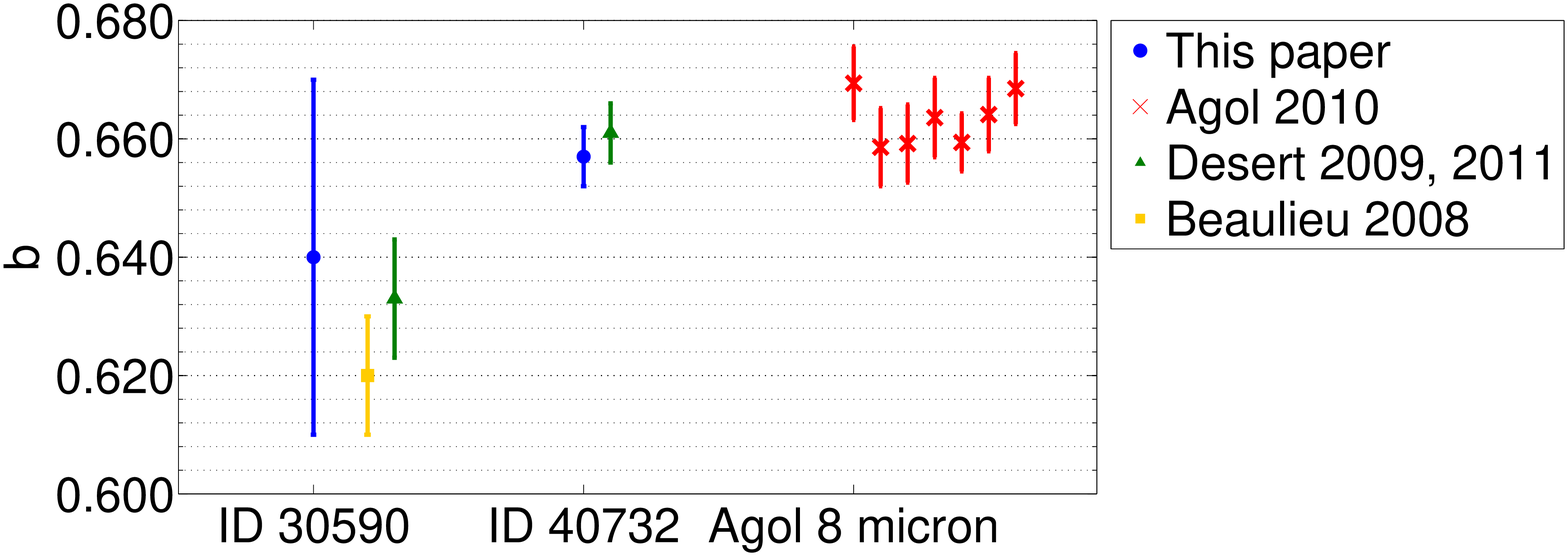}
\plotone{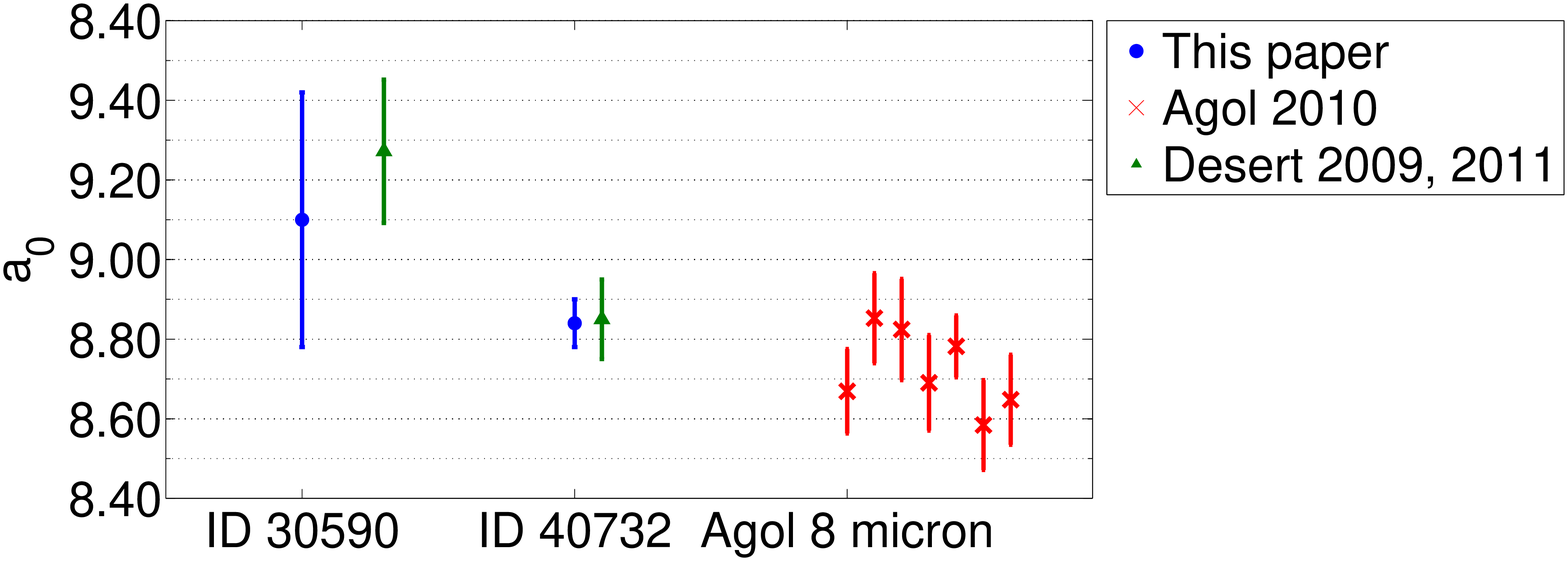}
\plotone{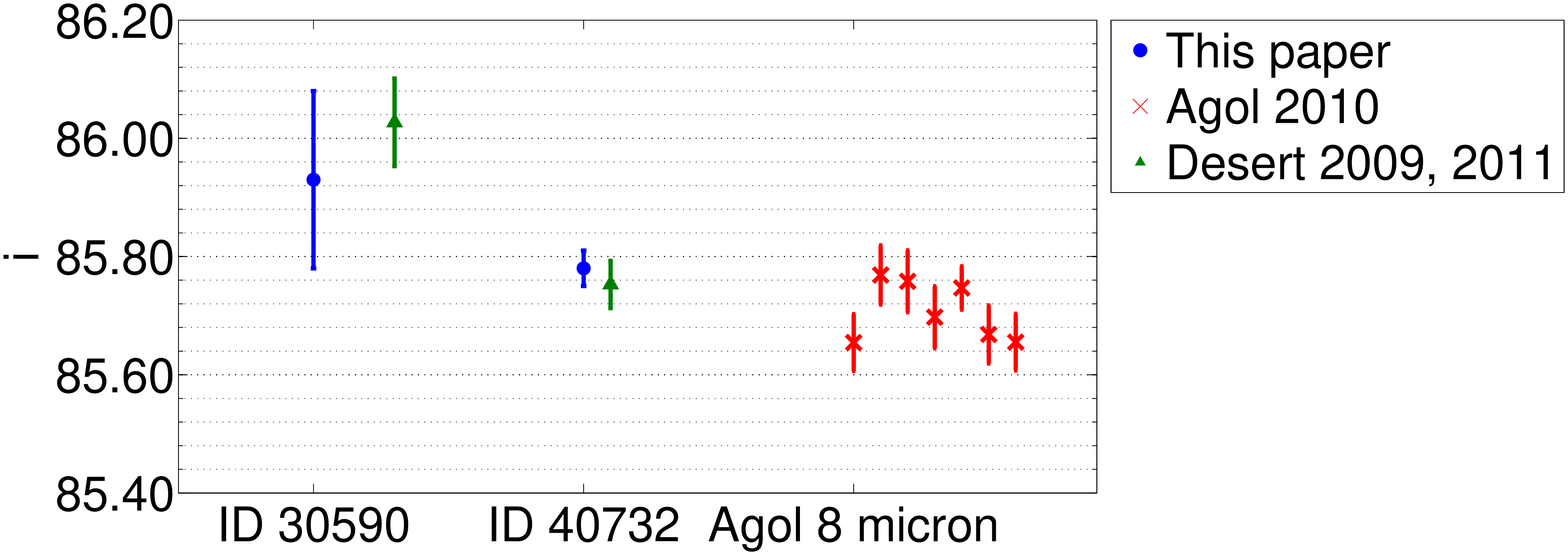}
\plotone{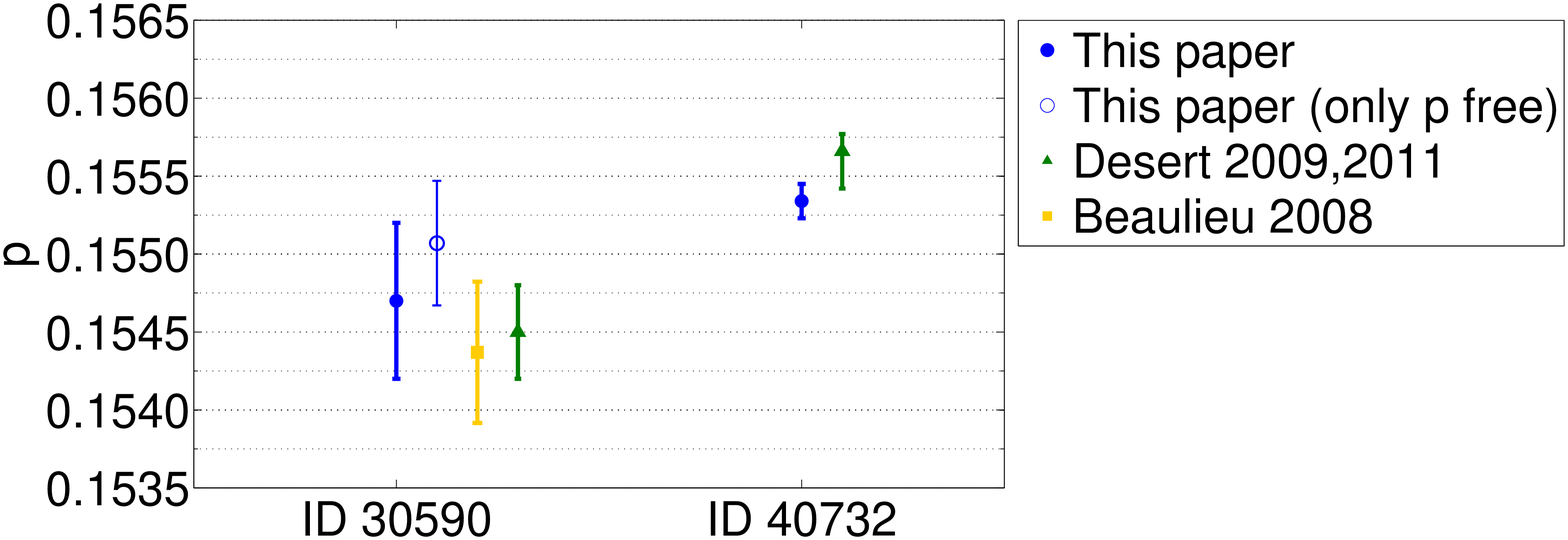}
\plotone{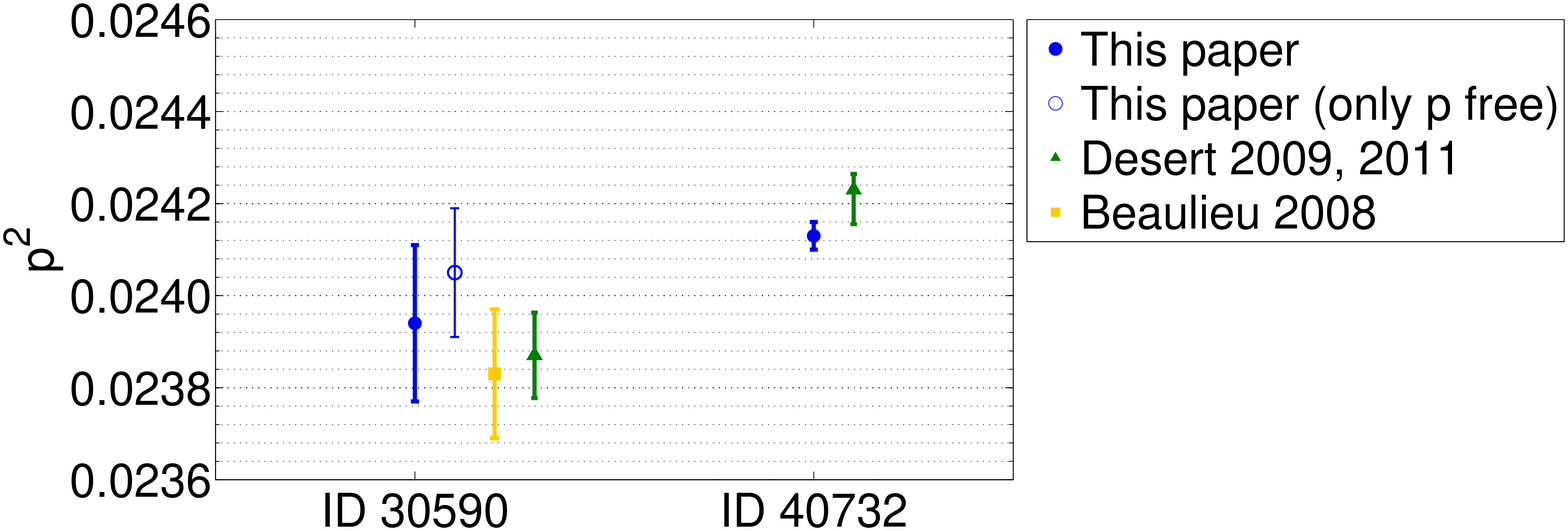}
\caption{From top to bottom: Comparisons of the parameters $b$, $a_0$, $i$, $p$, $p^2$,  obtained in this paper and in the others discussed here.
\label{fig11}}
\end{figure}

\cite{agol10} report a detailed study of seven primary transits and seven secondary eclipses of HD189733b, observed with Spitzer/IRAC at 8$\mu$m (channel 4 of IRAC). Their measured orbital parameters differ from ours by less than the joint 1-$\sigma$ uncertainties. Fig. \ref{fig11} includes a comparison of the parameters $a_0$, $i$, and $b$, obtained in this paper with their values. Given the number of primary transits and secondary eclipses they analysed, and the small impact of the limb darkening effect at 8$\mu$m, this is a robust confirmation of the validity of our results at 3.6$\mu$m. We suggest the use of these parameters for future observations at other wavelengths. \\
\cite{agol10} found variations in the transit depth with a range of $\sim 2 \times 10^{-4}$ on $p^2$. We could not detect such a difference between the two observations analysed at 3.6$\mu$m, as it is comparable with the first error bar.

\subsection{Comparison with previous analyses of the same observations}

Our results are consistent, at $1 \sigma$ level, with those of \cite{bea08}, \cite{des09} for ID 30590, and \cite{des11} for ID 40732. However, our results afford a substantial agreement (within 1$\sigma$) between the transit parameters determined from the two observations, while previous analyses by \cite{des09} and \cite{des11} claimed significant variations of all parameters (e.g., discrepancy $> 4\sigma$ for transit depths). \cite{des11} suggested stellar activity as a possible explanations for those differences. Our results do not support such conclusions, and we find that any possible stellar-activity variations are within the error bars.
Our error bars from the observation ID 40732 are of the same order (for transit depth) or even smaller (for orbital parameters) than in \cite{des11}, while those from the observation ID 30590 are larger by a factor $\sim 1.6$ with respect to the error bars in \cite{des09}. The factor $\sim 1.6$ comes from adding the ICA errors to the parameter error bars derived from the extracted signals. \cite{des09, des11} applied parametric corrections to detrend the transit signals from other disturbances, without attributing any uncertainties to the detrending processes. The fact that we obtained smaller error bars from the observation ID 40732, even including the contributions from the detrending process, indicates that, in that context, our blind extraction performed better than their parameterisation.
Orbital parameters determined by \cite{bea08} and \cite{des09} for observation ID 30590 are not consistent with those for observation ID 40732 obtained by \cite{des11}, the results presented here, or the 8$\mu$m observations by \cite{agol10}. Given that the second measurement was superior in quality, and given the agreement with observations at another wavelength, we conclude that the parameters presented in this paper are more robust than those reported by \cite{bea08}, or by \cite{des09} using the same data.

We note that \cite{bea08} used the same impact parameter at 3.6$\mu$m and 5.8$\mu$m, while \cite{des09} used similar, but not identical, values. Given the conclusions obtained in this paper about the orbital parameters, we suggest that the transit depth at 5.8$\mu$m be recalculated accordingly. A re-analysis of the observation at 5.8$\mu$m, then the differences between the transit depths at the two wavelengths, which were used to infer about the atmosphere of the planet, should not be strongly affected by this bias, at least in the first case. However, because their conclusions were controversial, a re-analysis of the observation at $5.8 \mu m$, with more precise orbital parameters and possibly non parametric technique, as done here, is needed. \\

\section{Conclusions}

We have introduced a blind signal-source separation method, based on ICA, to analyse photometric data of transiting exoplanets, with a high degree of objectivity; a novel aspect is the use of pixel-lightcurves, rather than multiple observations. \\
We have applied the method to a reanalysis of two Spitzer/IRAC datasets at 3.6$\mu$m, which previous analyses found to give discrepant results, and obtained consistent transit parameters from the two observations. \\
We suggest that the large scatter of results reported in the literature arises from:
\begin{itemize}
\item use of arbitrary parametric methods to detrend the transit signals, neglecting the relevant uncertainties;
\item correlations between parameters in the lightcurve fit.
\end{itemize}
We found, for observation ID 40732, values for the orbital parameters that are in excellent agreement with those found by \cite{agol10}, based on Spitzer/IRAC observations at $8 \mu m$.
By applying these values to observation ID 30590, we improved the accuracy of the inferred transit depth, and strengthened the consistency between the two observations.

\acknowledgements

G. Morello was partly funded by Erasmus (LLP), ``Borse di Studio finalizzate alla ricerca e Assegni finanziati da Programmi Comunitari, decreto 3505/2012'' of Universit\`a degli Studi di Palermo, Perren/Impact (CJ4M/CJ0T). G. Tinetti is a Royal Society URF. Part of this work was supported by STFC, and ASI-INAF agreement I/022/12/0.

\appendix

\section{Partial results}
\label{sec:app0}

\subsection{Observation ID 30590}

Tab. \ref{tab5} reports the best values of the parameters for the transit signals extracted from different arrays of pixels, the standard deviations of the residuals between the signals and the best transit models, and the standard deviations attributed to the ICA separation.\\
\begin{table}[!h]
\begin{center}
\caption{Best values of $p$, $a_0$, and $i$ for the transit signals extracted from different arrays of pixels, through method 2, considering all the independent components. Correspondents $\sigma_{0}$ (computed by the residuals between the signals and the best models, binned by 9 points), and $\sigma_{ICA}$. Derived total standard deviations of the parameters' distributions (observation ID 30590). \label{tab5}}
\begin{tabular}{cccccc}
\tableline\tableline
 & $3 \times 3$ & $5 \times 5$ & $7 \times 7$ & $9 \times 9$ & $11 \times 11$\\
\tableline
$p$ & $0.1549$ & $0.1547$ & $0.1546$ & $0.1547$ & $0.1547$\\
$a_0$ & $9.02$ & $9.05$ & $9.07$ & $9.06$ & $9.07$\\
$i$ & $85.90$ & $85.93$ & $85.95$ & $85.94$ & $85.95$\\
\tableline
$\sigma_{0}$ & $7.1 \cdot 10^{-4}$ & $6.5 \cdot 10^{-4}$ & $6.5 \cdot 10^{-4}$ & $6.5 \cdot 10^{-4}$ & $6.5 \cdot 10^{-4}$\\
$\sigma_{ICA}$ & $7.8 \cdot 10^{-4}$ & $7.6 \cdot 10^{-4}$ & $7.4 \cdot 10^{-4}$ & $8.2 \cdot 10^{-4}$ & $8.3 \cdot 10^{-4}$\\
\tableline
$\sigma_p$ & $0.00058$ & $0.00055$ & $0.00054$ & $0.00057$ & $0.00058$\\
$\sigma_{a_0}$ & $0.17$ & $0.16$ & $0.16$ & $0.17$ & $0.17$\\
$\sigma_i$ & $0.15$ & $0.15$ & $0.14$ & $0.15$ & $0.15$\\
\tableline
\end{tabular}
\end{center}
\end{table}

\begin{figure}
\epsscale{.80}
\plotone{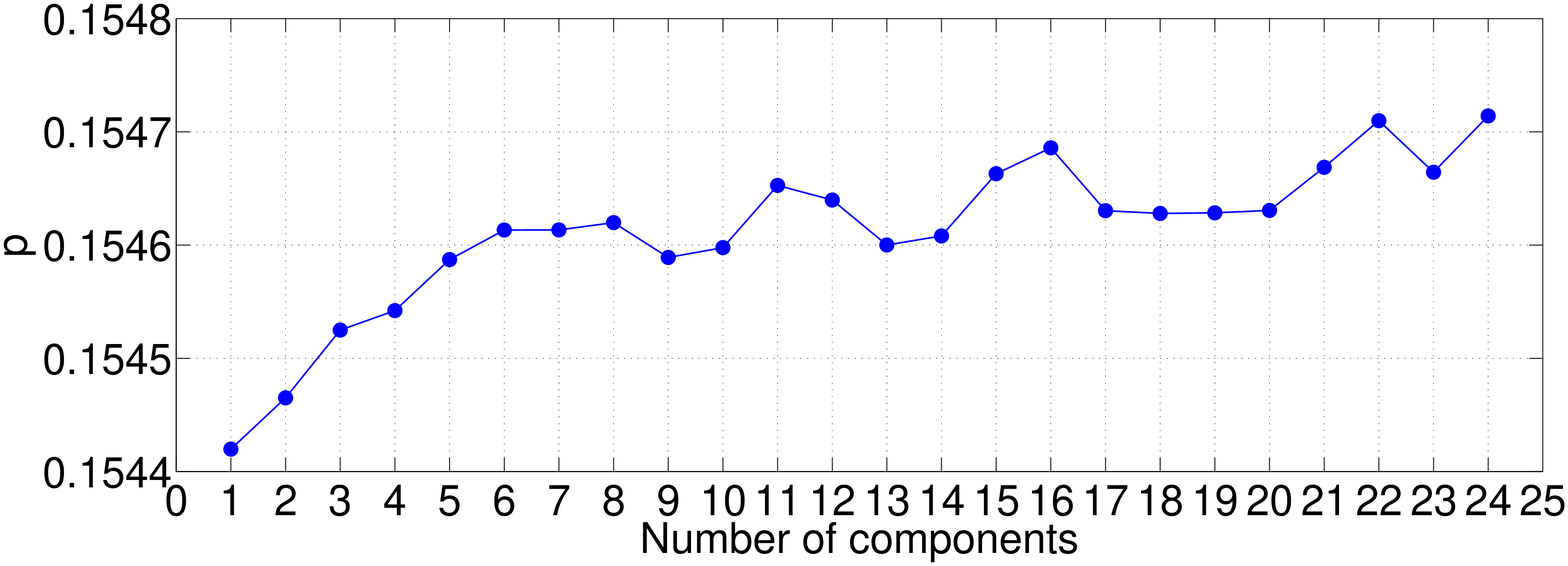}
\plotone{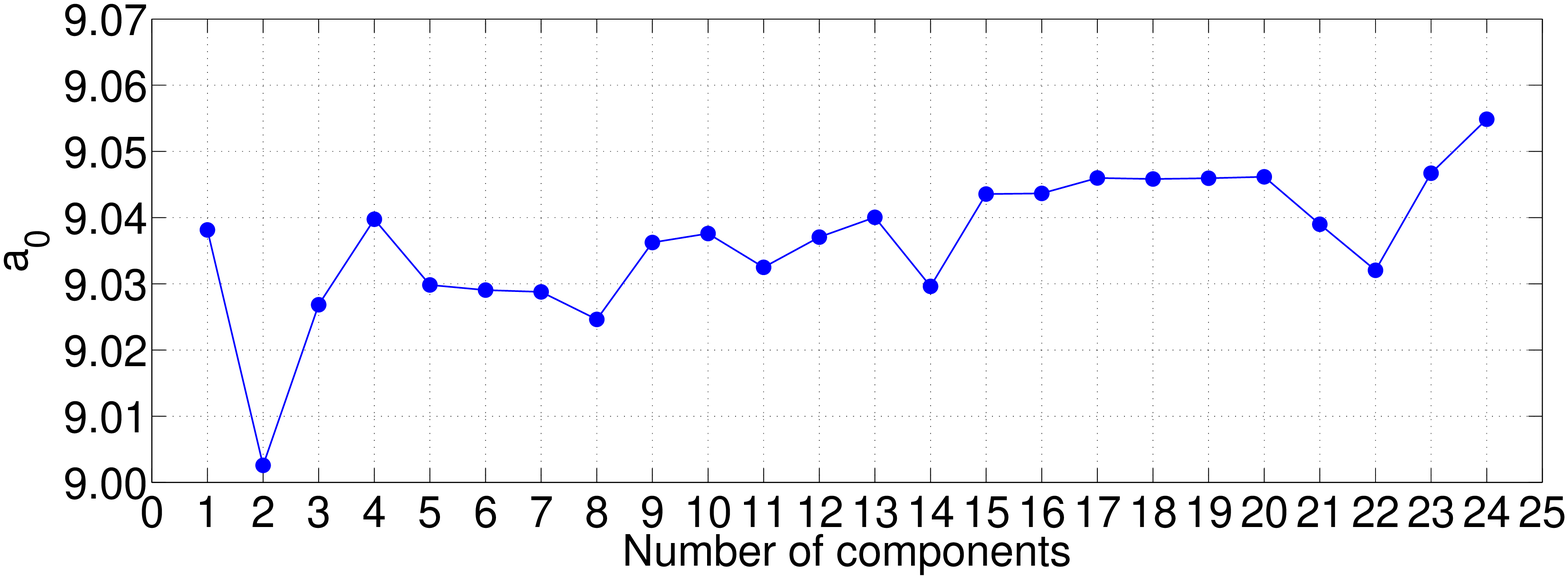}
\plotone{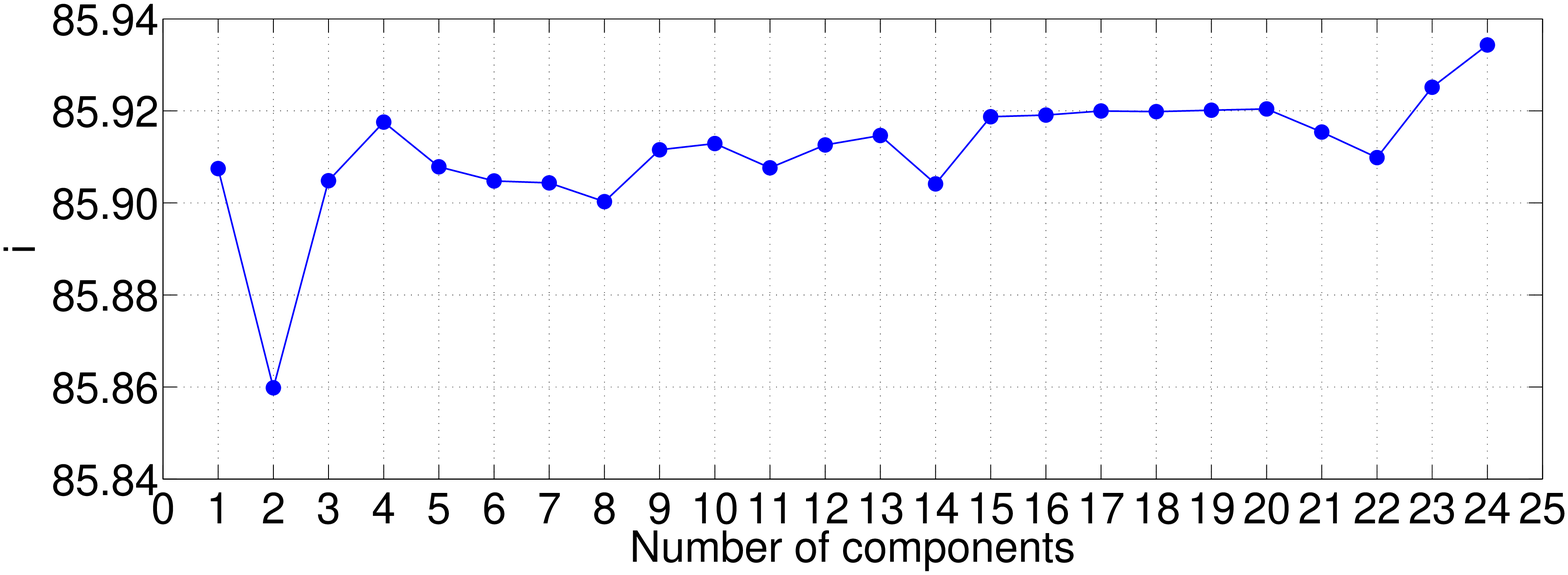}
\caption{From top to bottom: best values of the parameters $p$, $a_0$, and $i$, respectively, for the transit signals extracted removing the $n$ most significant components from the integral $5 \times 5$ lightcurve, binned by nine points (observation ID 30590). \label{fig12}}
\end{figure}

Fig. \ref{fig12} shows the best values of the parameters $p$, $a_0$, and $i$, respectively, for the transit signals extracted removing the $n$ most significant components from the integral $5 \times 5$ lightcurve, binned by nine points.

\subsection{Observation ID 40732}
Tab. \ref{tab6} reports the best values of the parameters for the transit signals extracted from different arrays of pixels, the standard deviations of the residuals between the signals and the best transit models, and the standard deviations attributed to the ICA separation.
\begin{table}[!h]
\begin{center}
\caption{Best values of $p$, $a_0$, and $i$ for the transit signals extracted from different arrays of pixels, through method 2, considering all the independent components, binned by 9 points. Correspondents $\sigma_{0}$ (computed by the residuals between the signals and the best models), and $\sigma_{ICA}$. Derived total standard deviations of the parameters' distributions (observation ID 40732). \label{tab6}}
\begin{tabular}{cccccc}
\tableline\tableline
 & $3 \times 3$ & $5 \times 5$ & $7 \times 7$ & $9 \times 9$ & $11 \times 11$\\
\tableline
$p$ & $0.15546$ & $0.15534$ & $0.15533$ & $0.15533$ & $0.15528$\\
$a_0$ & $8.93$ & $8.92$ & $8.93$ & $8.93$ & $8.94$\\
$i$ & $85.79$ & $85.78$ & $85.79$ & $85.79$ & $85.79$\\
\tableline
$\sigma_{0}$ & $1.62 \cdot 10^{-4}$ & $1.46 \cdot 10^{-4}$ & $1.45 \cdot 10^{-4}$ & $1.45 \cdot 10^{-4}$ & $1.41 \cdot 10^{-4}$\\
$\sigma_{ICA}$ & $1.70 \cdot 10^{-4}$ & $1.45 \cdot 10^{-4}$ & $1.60 \cdot 10^{-4}$ & $1.53 \cdot 10^{-4}$ & $1.65 \cdot 10^{-4}$\\
\tableline
$\sigma_p$ & $0.00013$ & $0.00011$ & $0.00012$ & $0.00011$ & $0.00012$\\
$\sigma_{a_0}$ & $0.03$ & $0.03$ & $0.03$ & $0.03$ & $0.03$\\
$\sigma_i$ & $0.03$ & $0.03$ & $0.03$ & $0.03$ & $0.03$\\
\tableline
\end{tabular}
\end{center}
\end{table}

\begin{figure}[!h]
\epsscale{.80}
\plotone{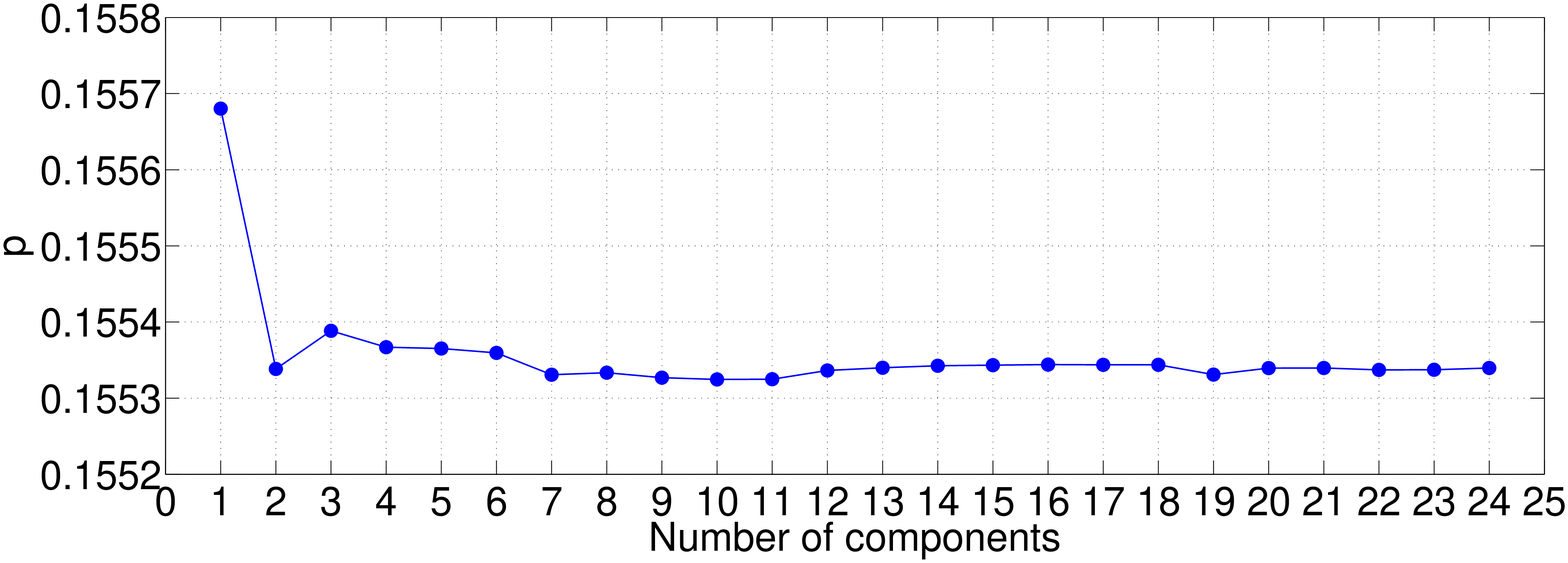}
\plotone{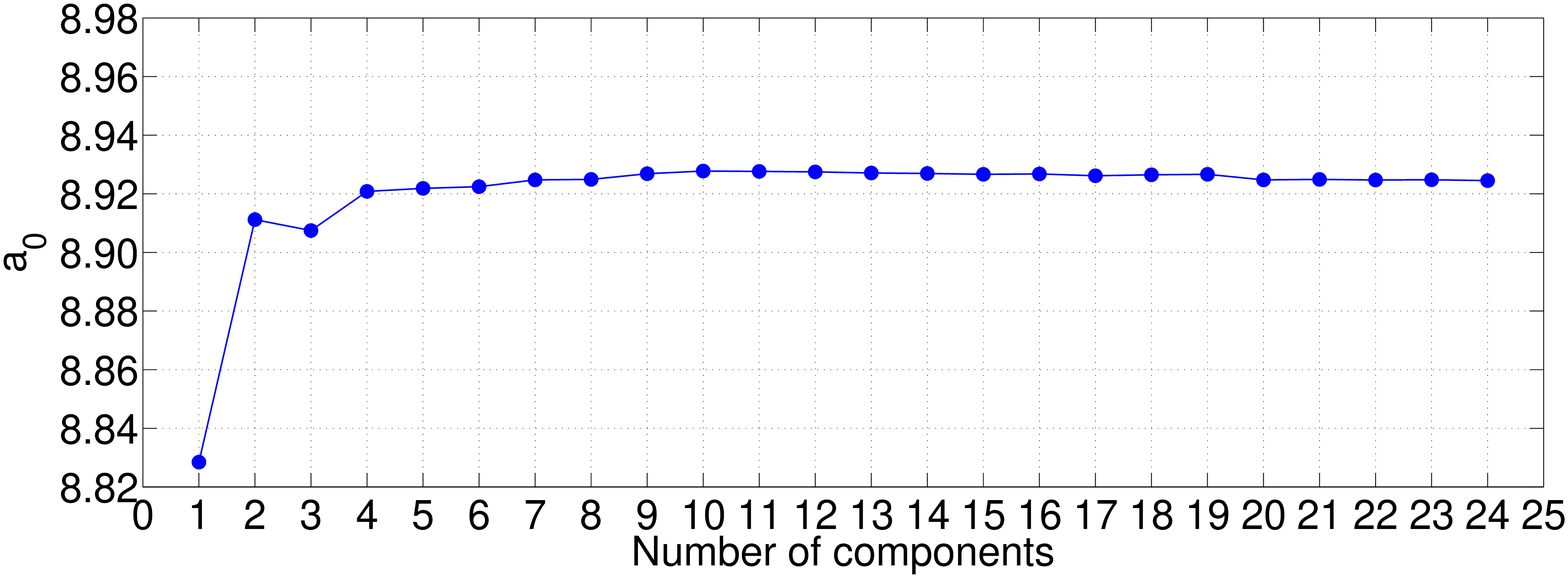}
\plotone{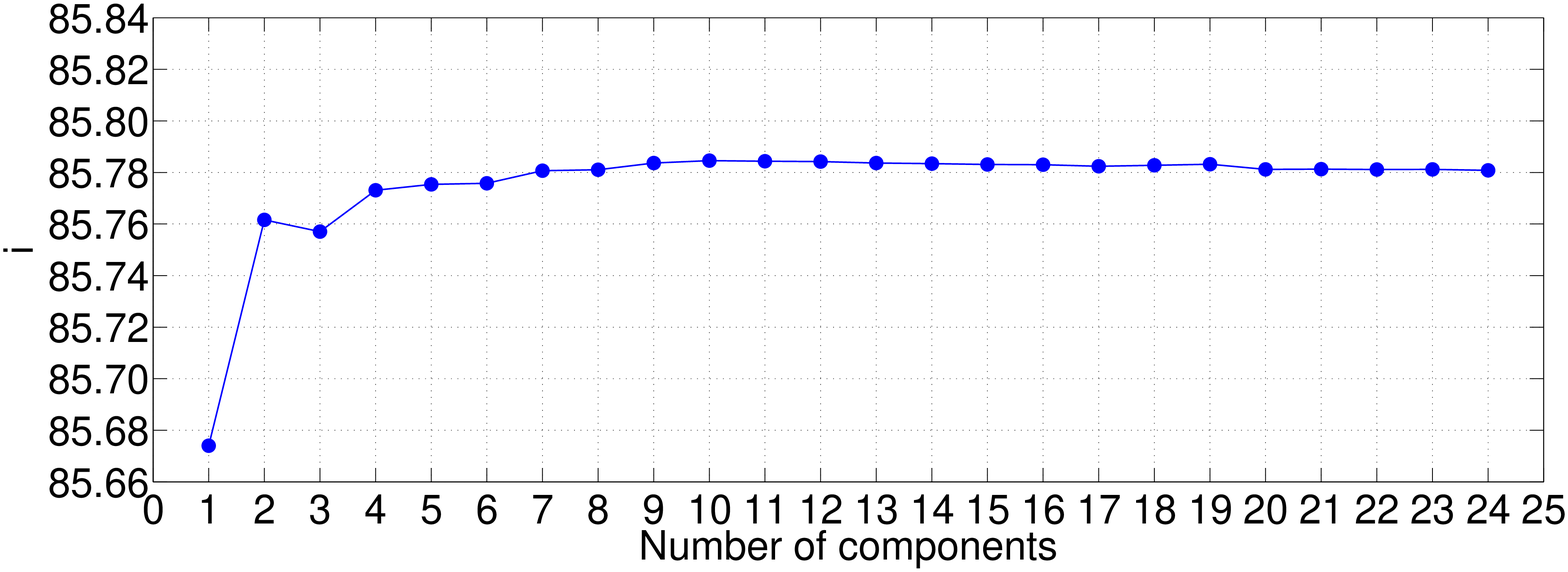}
\caption{From top to bottom: best values of the parameters $p$, $a_0$, and $i$, respectively, for the transit signals extracted removing the other-components-models, with the $n$ most significant components, from the integral $5 \times 5$ lightcurve, binned by nine points (observation ID 40732). \label{fig13}}
\end{figure}

Fig. \ref{fig13} reports the best values of the parameters $p$, $a_0$, and $i$, respectively, for the transit signals extracted removing the $n$ most significant components from the integral $5 \times 5$ lightcurve, binned by nine points.

\section{Subdatasets}
\label{sec:app1}

An important test to verify the robustness of the analyses is to apply the same techniques to subdatasets. They clearly share the same phenomena, but recorded for different time intervals, largely overlapping. If the technique is able to separate the source components, the detrended transit signals from different subdatasets should be essentially equivalent, otherwise there is a problem with at least one of them. A critical factor could be the time length of a subdataset compared to the timescales of the source signals; for this reason, the separation performed using longer subdatasets or the whole dataset, might be more reliable, unless they strengthen some trends or introduce bad data, for example if they are not well calibrated, or affected by spurious events. \\

\subsection{Observation ID 30590}

\begin{figure}[!h]
\epsscale{.80}
\plotone{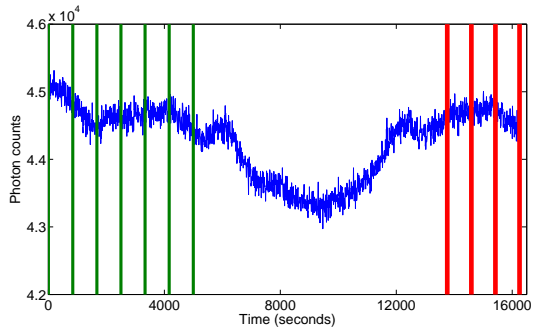}
\caption{The integral lightcurve from the $5 \times 5$ array. The green vertical lines indicate the different start points considered; the red vertical lines indicate the end points. (observation ID 30590) \label{fig14}}
\end{figure}

We considered twenty-eight subdatasets, obtained combining seven different starting and four ending times, disposed with regular cadence of $\sim$14 minutes (see Fig. \ref{fig14}).

\begin{figure}[!h]
\epsscale{.80}
\plotone{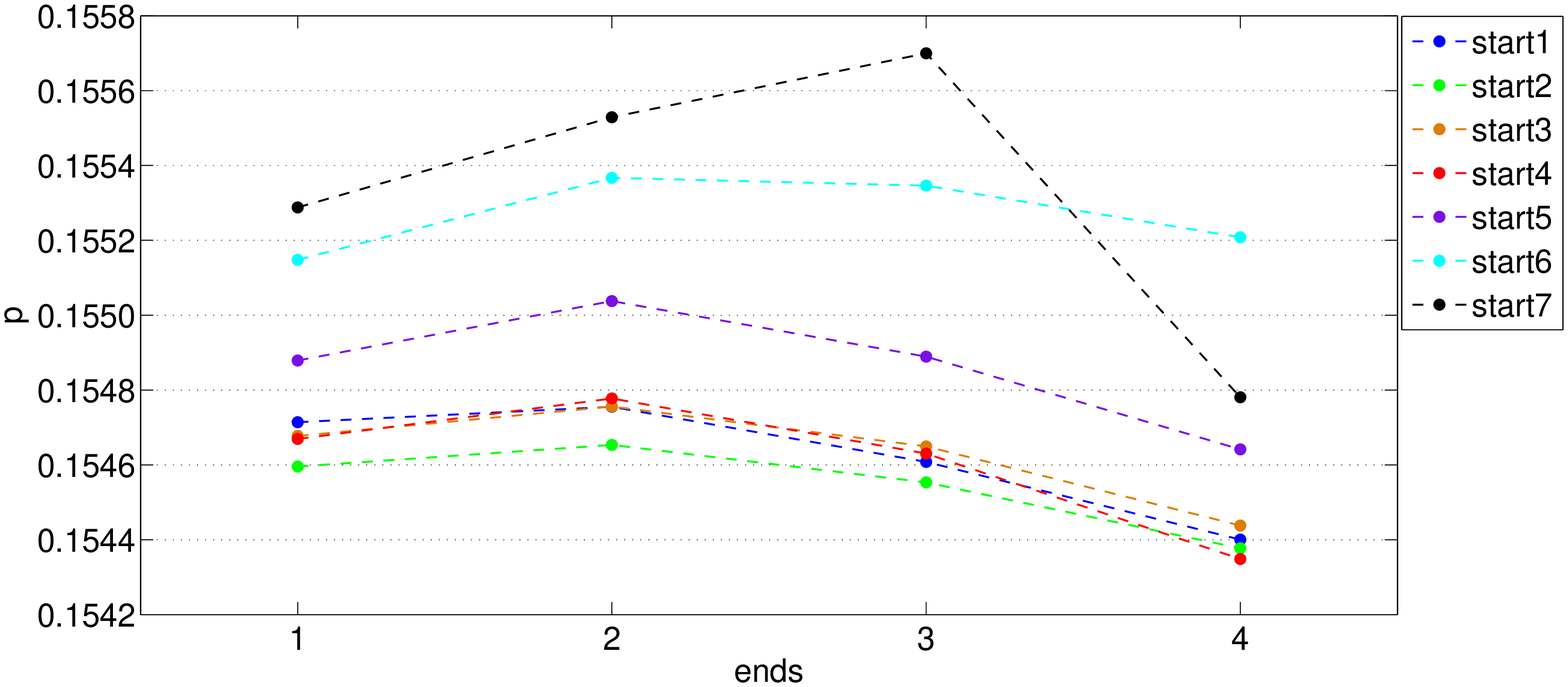}
\plotone{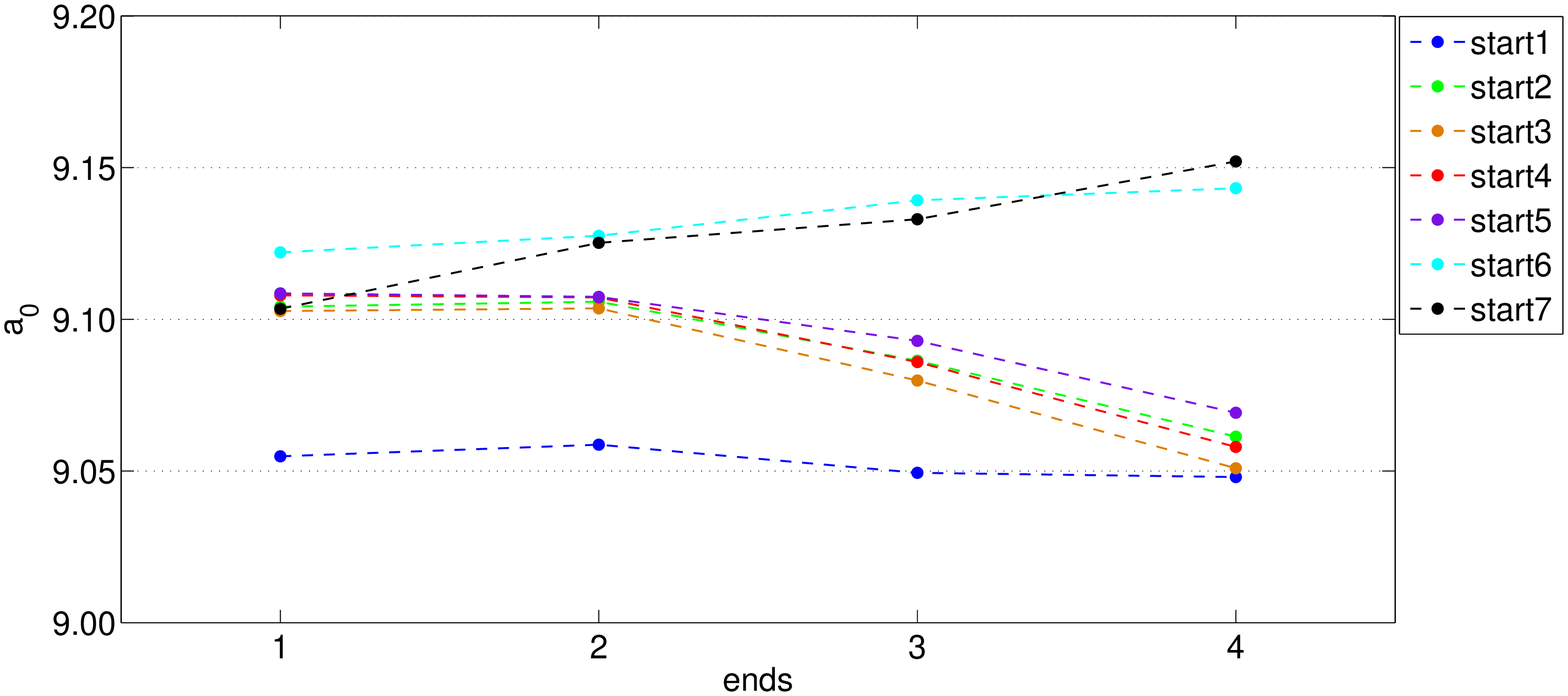}
\plotone{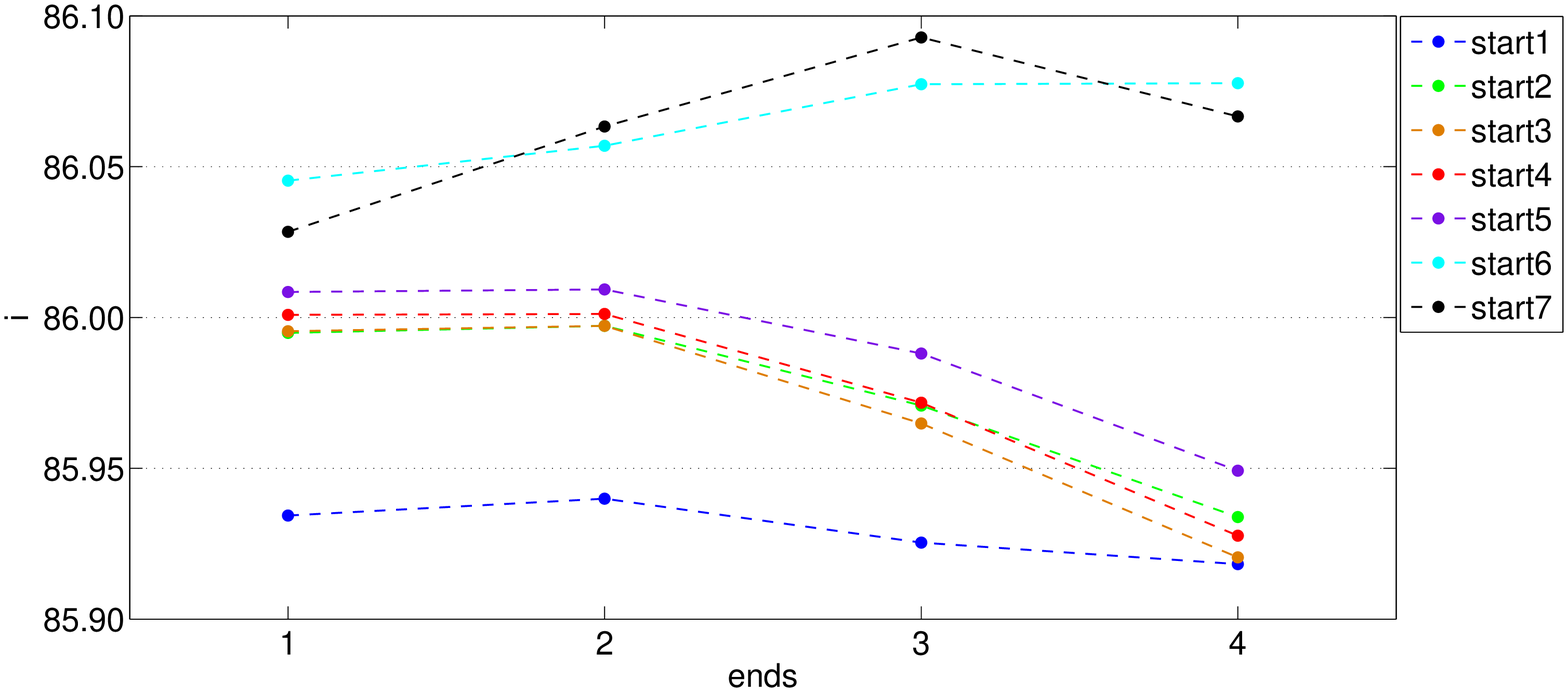}
\caption{From top to bottom: Best values of the parameters $p$, $a_0$, and $i$, respectively, for the transit signals obtained through method 2, from different subdatasets. They are extracted using the $5 \times 5$ array, by removing all the independent components from the integral lightcurve. The curve were binned by nine points, before performing the fits. Different colours are used depending on the starts, indexed from earlier to later with increasing integers: blue, start 1, green, start 2, ecru, start 3, red, start 4, purple, start 5, cyan, start 6, black, start 7. Index from 1 to 4 on the horizontal axis indicate different ends, from later to earlier (observation ID 30590). \label{fig15}}
\end{figure}

As before, we used the $5 \times 5$ array, and we applied method 2, by removing all the independent components from the integral lightcurve.  Fig.  \ref{fig15} shows the best values of the parameters $p$, $a_0$, and $i$, estimated using each subdataset. \\
We can point out some correlations between the best values and both the start and the end points of the subdatasets. The overall scatters are compatible with the ranges determined before.  Tab. \ref{tab7} reports the estimated ranges of the parameters with the scatters observed by the subdatasets, either by including or by rejecting the two shortest subdatasets.
\begin{table}[!h]
\begin{center}
\caption{Best values and error bars of $p$, $a_0$, and $i$, overall scatters observed by using different subdatasets, and by rejecting the two shortest ones. (observation ID 30590) \label{tab7}}
\begin{tabular}{cccc}
\tableline \tableline
Parameters & Estimated values & Overall scatters by subdatasets & With rejections \\
\tableline
$p$ & $0.1547 \pm 0.0005$ & $0.1543 \div 0.1557$ & $0.1543 \div 0.1550$\\
$a_0$ & $9.05 \pm 0.16$ & $9.05 \div 9.15$ & $9.05 \div 9.11$\\
$i$ & $85.93 \pm 0.15$ & $85.92 \div 86.09$ & $85.92 \div 86.01$\\
\tableline
\end{tabular}
\end{center}
\end{table}

\subsection{Observation ID 40732}

\begin{figure}[!h]
\epsscale{.80}
\plotone{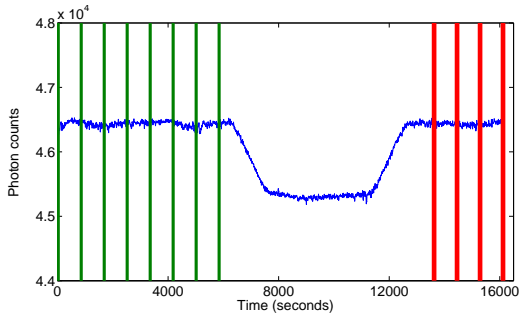}
\caption{The integral lightcurve from the $5 \times 5$ array. The green vertical lines indicate the different start points considered; the red vertical lines indicate the end points (observation ID 40732). \label{fig16}}
\end{figure}

We considered thirty-two subdatasets, obtained combining eight different starting times and four ending times, disposed with regular cadence of $\sim$14 minutes (see Fig. \ref{fig16}).

\begin{figure}[!h]
\epsscale{.80}
\plotone{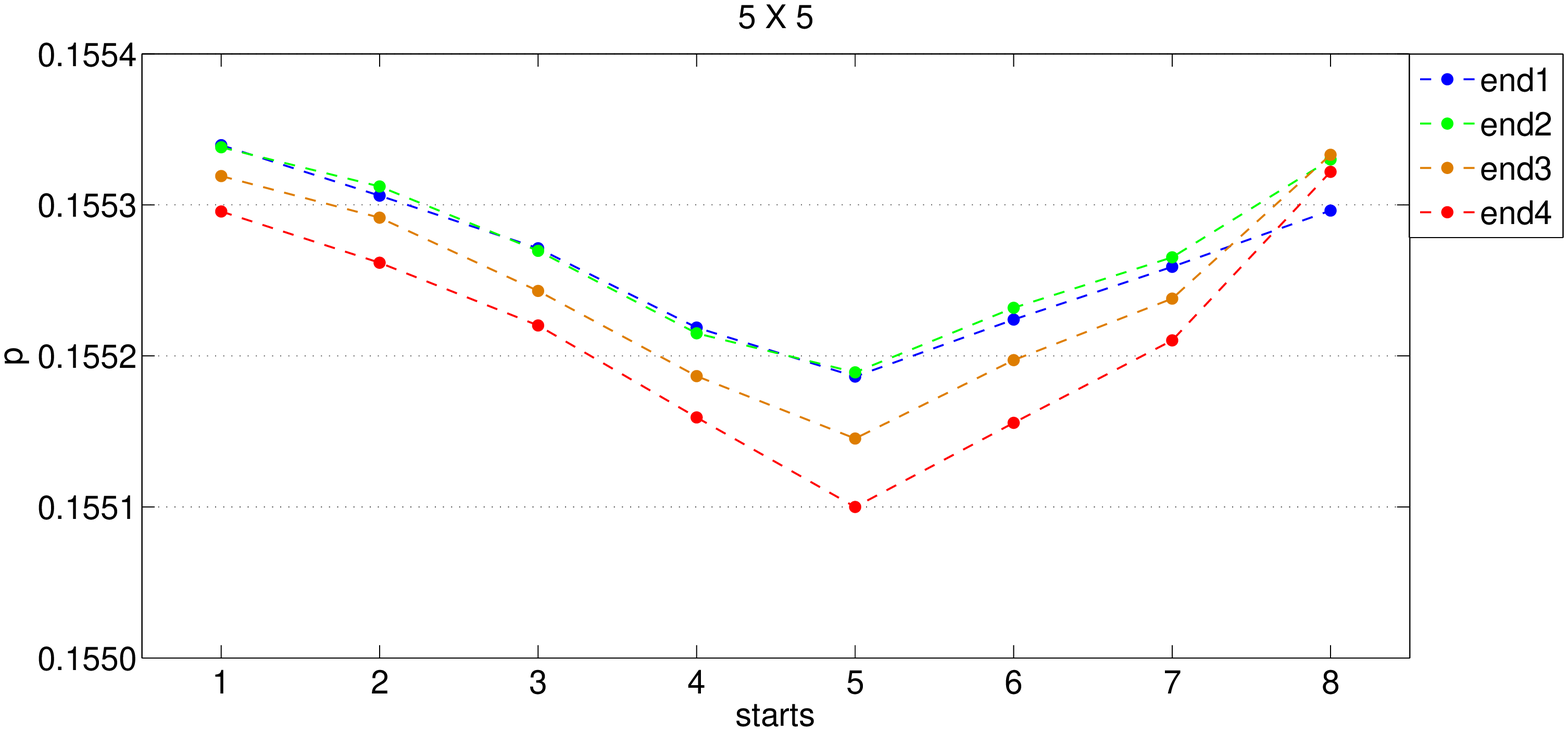}
\plotone{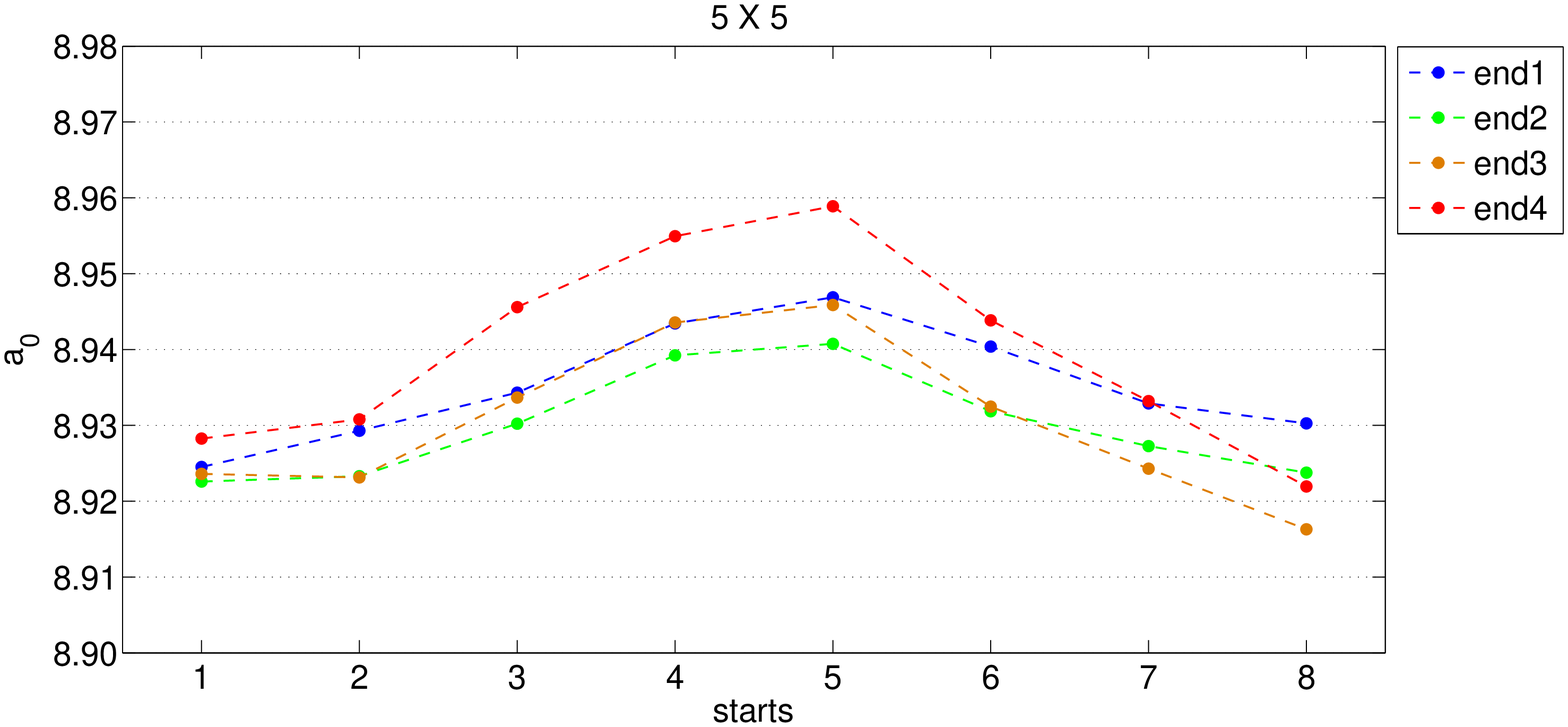}
\plotone{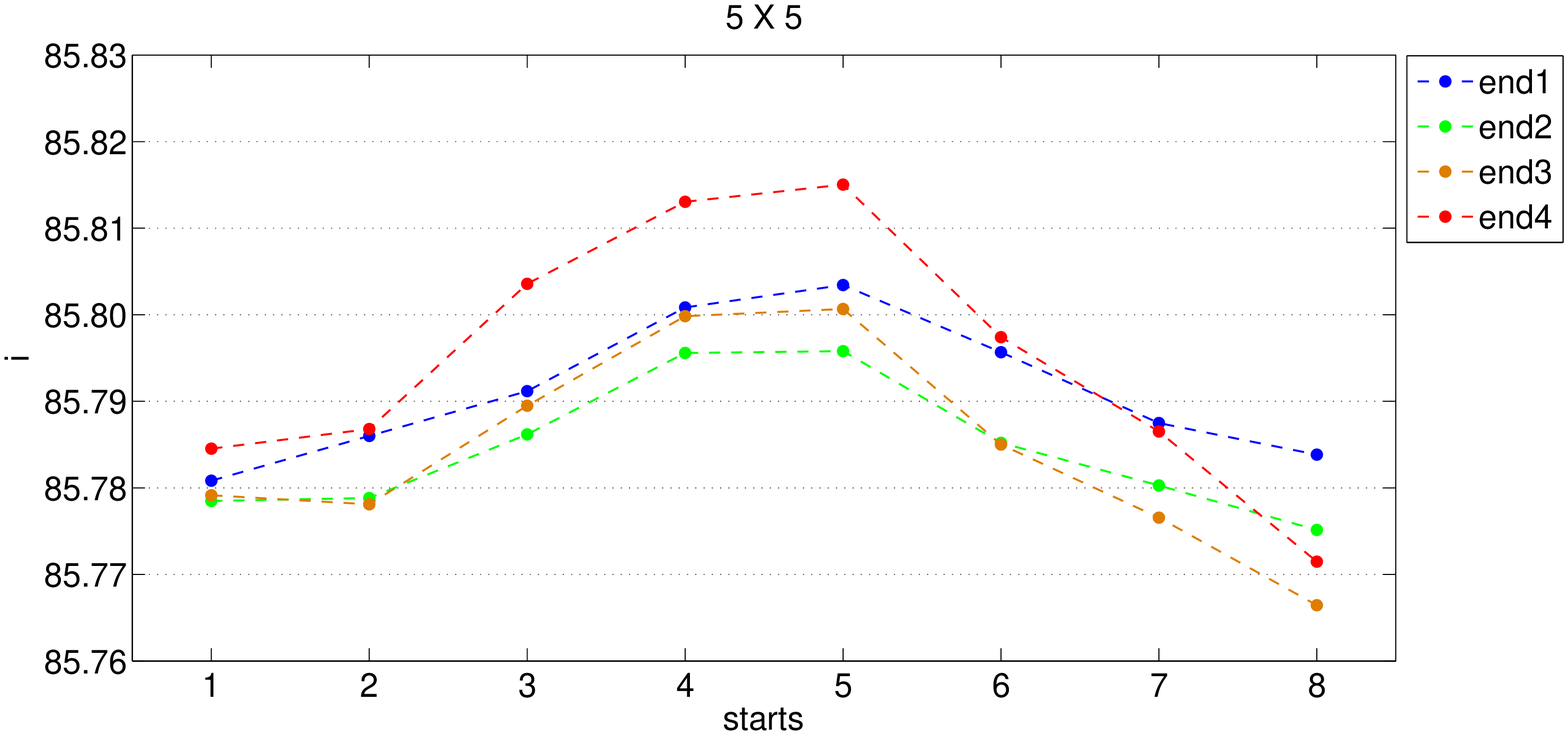}
\caption{From top to bottom: Best values of the parameters $p$, $a_0$, and $i$, respectively, for the transit signals obtained through method 2, from different subdatasets. They are extracted using the $5 \times 5$ array, by removing all the independent components from the integral lightcurve. The curve were binned by nine points, before performing the fits. Different colours are used depending on the ends, indexed from later to earlier with increasing integers: blue, end 1, green, end 2, ecru, end 3, red, end 4. Index from 1 to 8 on the horizontal axis indicate different starts, from later to earlier (observation ID 40732). \label{fig17}}
\end{figure}

As usual, we used the $5 \times 5$ array, and we applied method 2, and removed all the independent components from the integral lightcurve. Fig. \ref{fig17} shows the best values of the parameters $p$, $a_0$, and $i$, estimated using each subdataset.

Again, there are some correlations between the best values and the extremes of the subdatasets, but the overall scatters are compatible with the ranges previously estimated. Tab. \ref{tab8} reports the estimated ranges of the parameters with the scatters observed by the subdatasets: \\
\begin{table}[!h]
\begin{center}
\caption{Best values and error bars of $p$, $a_0$, and $i$, overall scatters observed by using different subdatasets, and by rejecting the two shortest ones (observation ID 40732). \label{tab8}}
\begin{tabular}{ccc}
\tableline \tableline
Parameters & Estimated values & Overall scatters by subdatasets \\
\tableline
$p$ & $0.15534 \pm 0.00011$ & $0.15510 \div 0.15534$\\
$a_0$ & $8.92 \pm 0.03$ & $8.92 \div 8.96$\\
$i$ & $85.78 \pm 0.03$ & $85.77 \div 85.82$\\
\tableline
\end{tabular}
\end{center}
\end{table}

\section{Method 1: direct identification of the transit component}
\label{sec:app2}

Tab. \ref{tab9} reports the results obtained by applying method 1 and method 2 on both observations, using the whole datasets, and the $5 \times 5$ arrays.
\begin{table}[!b]
\begin{center}
\caption{Estimated best values and error bars of $p$, $a_0$, $i$, $p^2$, $b$, and $T$ by applying method 1 and method 2 (both observations). \label{tab9}}
\begin{tabular}{ccc}
\tableline\tableline
ID 30590 & Method 1 & Method 2\\
\tableline
$p$ & $0.1547 \pm 0.0019$ & $0.1547 \pm 0.0005$\\
$a_0$ & $9.1 \pm 0.5$ & $9.05 \pm 0.16$\\
$i$ & $85.9 \pm 0.5$ & $85.93 \pm 0.15$\\
$p^2$ & $0.0239 \pm 0.0006$ & $0.02394 \pm 0.00017$\\
$b$ & $0.64 \pm 0.11$ & $0.64 \pm 0.03$\\
$T$ & $5160 \pm 900 \ s$ & $5170 \pm 200 \ s$\\
\tableline
\end{tabular}
\end{center}
\begin{center}
\begin{tabular}{ccc}
\tableline\tableline
ID 40732 & Method 1 & Method 2\\
\tableline
$p$ & $0.1553 \pm 0.0004$ & $0.15534 \pm 0.00011$\\
$a_0$ & $8.96 \pm 0.10$ & $8.92 \pm 0.03$\\
$i$ & $85.81 \pm 0.11$ & $85.78 \pm 0.03$\\
$p^2$ & $0.02413 \pm 0.00012$ & $0.02413 \pm 0.00003$\\
$b$ & $0.654 \pm 0.019$ & $0.657 \pm 0.005$\\
$T$ & $5156 \pm 124 \ s$ & $5157 \pm 34 \ s$\\
\tableline
\end{tabular}
\end{center}
\end{table}
It is straightforward to note that the best values are almost coincident, but the uncertainties derived with method 1 are larger by a factor $\sim 3 \div 4$. The differences are due to the ICA contributions to the error bars. \\

\begin{figure}[!h]
\epsscale{.80}
\plotone{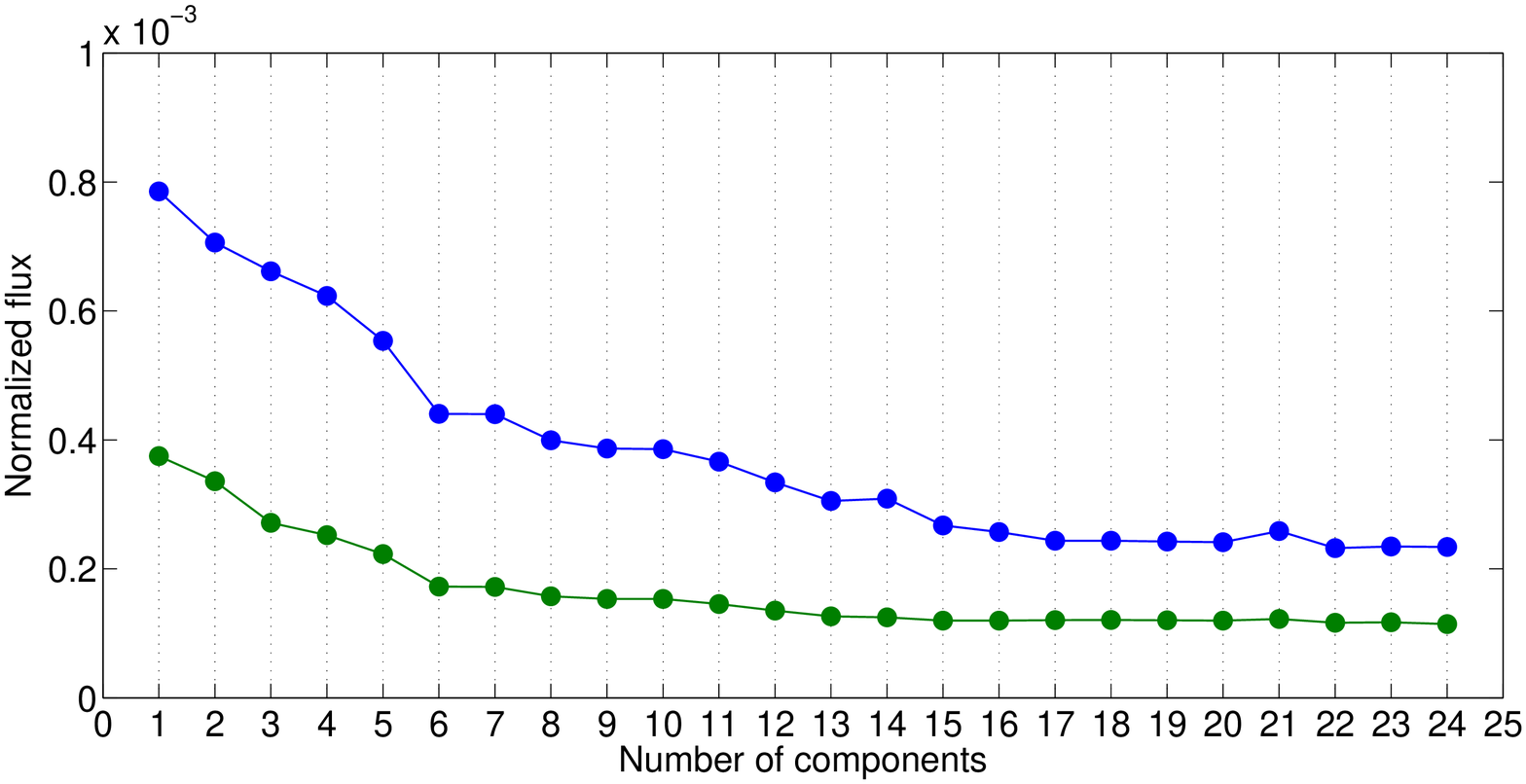}
\plotone{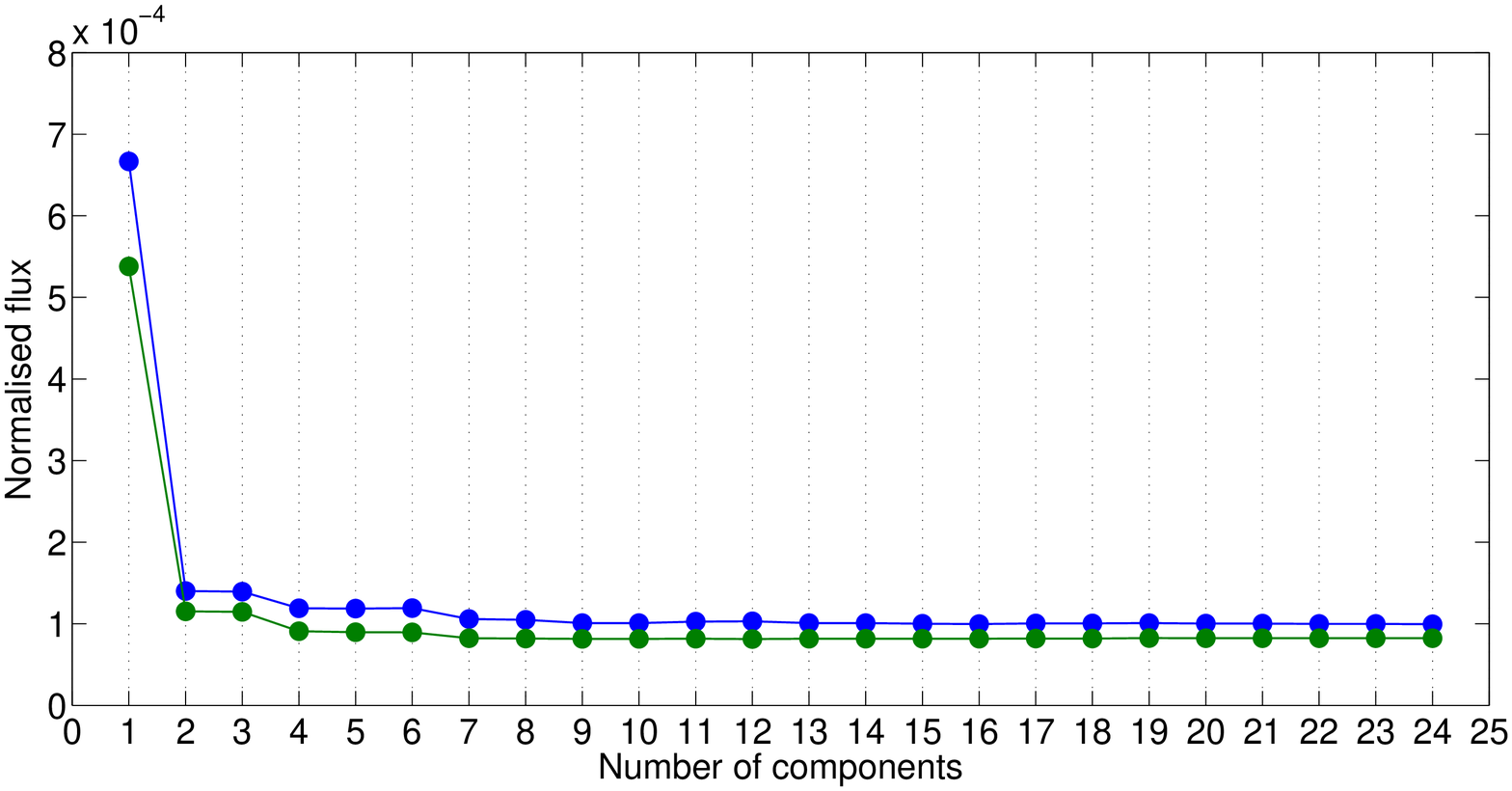}
\caption{Top: Observation ID 30590; blue, mean quadratic deviations between the transit signals estimated through method 2, with the $n$ most important components, and the one estimated through method 1, using the $5 \times 5$ array; green, the same, considering the binned signals.  Bottom: The same for observation ID 40732. \label{fig18}}
\end{figure}

We also observed that, in these cases, the transit signals estimated with method 2 tend to the ones obtained by method 1, when increasing the number of non-transit-components removed; this is shown in Fig. \ref{fig18}.

However, the larger error bars provided by the ICA terms are justyfied by the scatters obtained by using different arrays of pixels and different subdatasets. We do not report the results in detail, but we summarise the main facts observed:
\begin{itemize}
\item In some cases, the transit component is clearly corrupted, discouraging a quantitative analysis;
\item The scatters of the transit parameters obtained by using different subdatasets are comparable with the error bars estimated (the arrays of pixels play a minor role, but more important than if using method 2);
\item For longer subdatasets, which are expected to allow better extractions of the independent components, the results obtained with methods 1 and 2 tend to agree.
\end{itemize}





\end{document}